\documentclass[fleqn,usenatbib]{mnras}

\usepackage{newtxtext,newtxmath}
\usepackage[T1]{fontenc}
\usepackage{ae,aecompl}
\usepackage{graphicx}	% Including figure files
\usepackage{subfigure}
\usepackage{amsmath}	% Advanced maths commands
\usepackage{amssymb}	% Extra maths symbols
\usepackage{comment}
\usepackage{natbib}
\newcommand{\q}{\mathbf{q}}
\newcommand{\x}{\mathbf{x}}
\newcommand{\p}{\boldsymbol{\Psi}}
\newcommand{\pa}{\partial}

\title{Evolving Ultralight Scalars into Non-Linearity with Lagrangian Perturbation Theory}
\author[Lagu\"e et al.]{Alex Lagu\"e$^{1,2,3}$,\thanks{E-mail: lague@cita.utoronto.ca}
J. Richard Bond$^{3}$,
Ren\'ee Hlo\v zek$^{1,2}$,
\newauthor
David J. E. Marsh$^{4}$,
and Laurin S\"oding$^{5}$
\\
$^{1}$Department of Astronomy \& Astrophysics, University of Toronto, 50 St. George St., Toronto, ON, M5S 3H4, Canada\\
$^{2}$Dunlap Institute for Astronomy and Astrophysics, University of Toronto, 50 St. George St., Toronto, ON, M5S 3H4, Canada\\
$^{3}$Canadian Institute for Theoretical Astrophysics, University of Toronto, 60 St. George St., Toronto, ON, M5S 3H8, Canada\\
$^{4}$Institut fur Astrophysik, Georg-Agust Universitat, Friedrich-Hund-Platz 1, D-37077 Gottingen, Germany\\
$^{5}$Department of Physics and Astronomy, Universit\"at Heidelberg, 69120 Heidelberg, Germany
}

\date{Accepted XXX. Received YYY; in original form ZZZ}

\pubyear{2020}

\begin{document}
\label{firstpage}
\pagerange{\pageref{firstpage}--\pageref{lastpage}}
\maketitle

% Abstract of the paper
\begin{abstract}
Many models of high energy physics suggest that the cosmological dark sector consists of not just one, but a spectrum of ultralight scalar particles with logarithmically distributed masses. To study the potential signatures of low concentrations of ultralight axion (also known as fuzzy) dark matter, we modify Lagrangian perturbation theory (LPT) by distinguishing between trajectories of different dark matter species. We further adapt LPT to include the effects of a quantum pressure, which is necessary to generate correct initial conditions for ultralight axion simulations. Based on LPT, our modified scheme is extremely efficient on large scales and it can be extended to an arbitrary number of particle species at very little computational cost. This allows for computation of self-consistent initial conditions in mixed dark matter models. Additionally, we find that shell-crossing is delayed for ultralight particles and that the deformation tensor extracted from LPT can be used to identify the range of redshifts and scales for which the Madelung formalism of fuzzy dark matter is a reliable approximation.
\end{abstract}

% Select between one and six entries from the list of approved keywords.
% Don't make up new ones.

%%%%%%%%%%%%%%%%%%%%%%%%%%%%%%%%%%%
%%%%%%%%%%%%%%%%%%%%%%%%%%%%%%%%%%%

\section{Introduction}
\begin{figure*}
    \centering
    \includegraphics[width=\linewidth]{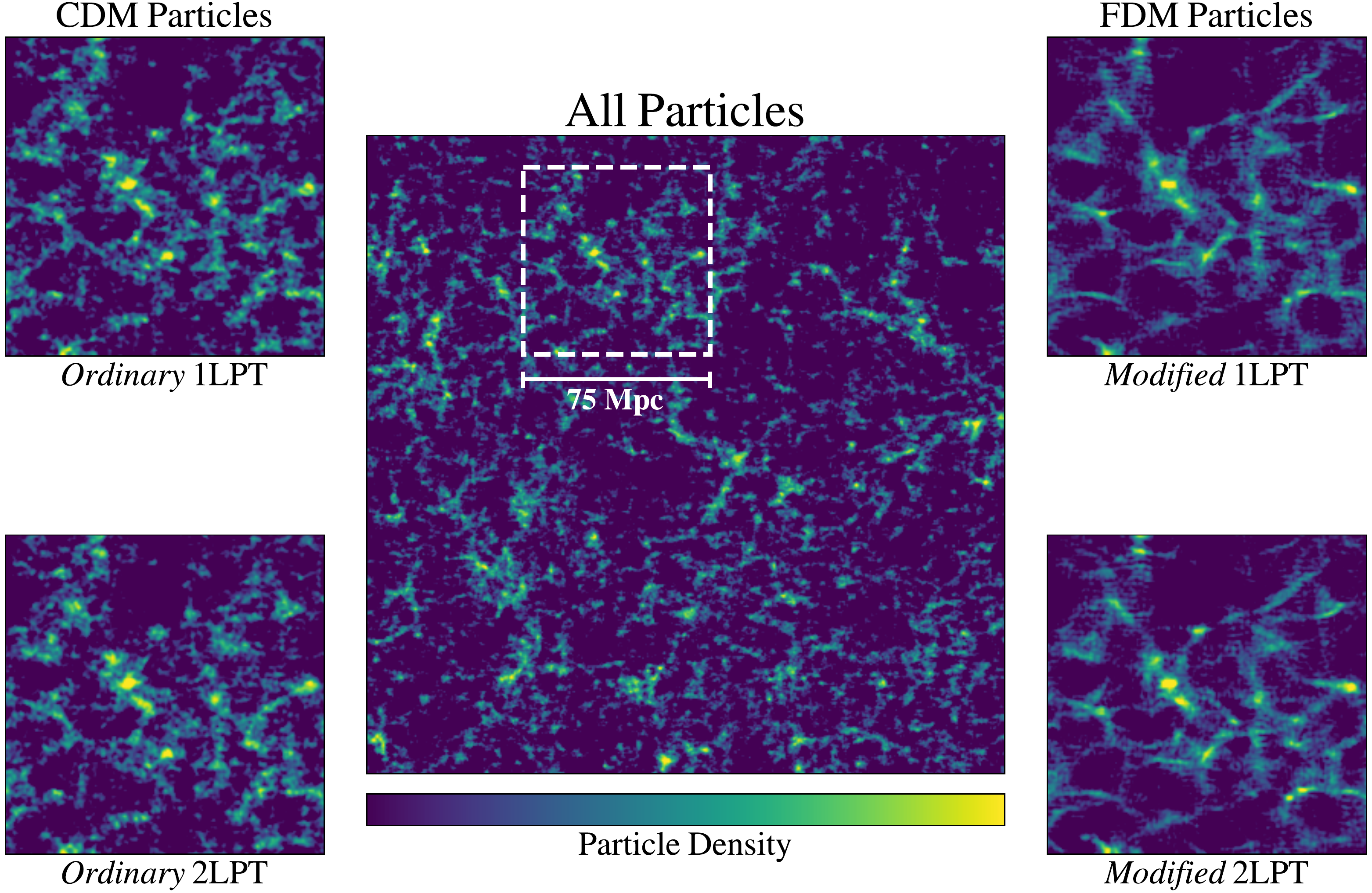}
    \caption{Composite particle number density of large-scale structure in a \emph{mixed} DM model composed at 90\% of CDM particles (left panel) and 10\% of FDM particles of mass $10^{-26}$ eV (right panel). Both are subject to the \emph{same} gravitational potential. The motion and clustering of test particles is computed using a modified first or second order Lagrangian perturbation theory (1LPT/2LPT) approach in a comoving (256 Mpc)${}^3$ simulation box.
    }
    \label{fig:mirror}
\end{figure*}
A light boson as a dark matter particle has been suggested as a solution to the small scale problems of $\Lambda$CDM (\citealt{Moore1998ResolvingHalos,Klypin1999WhereSatellites,Peebles2000FluidMatter,Hu2000FuzzyParticles,DeBlok2010TheProblem,Marsh2014AMatter,Bullock2017Small-ScaleParadigm}). A candidate for such particles are the ultralight axions (ULAs) which have a mass of $10^{-33}\: \mathrm{eV}\lesssim m \lesssim 10^{-21}\:\mathrm{eV}$. At masses on this scale, the particles have a de Broglie wavelength of astrophysical scale $\lambda_\mathrm{dB} \gtrsim 1\: \mathrm{kpc}$ (\citealt{Hu2000FuzzyParticles,Hui2017UltralightMatter}) and, as an example, a particle with mass $m\sim 5\times 10^{-24}$ eV would have a de Broglie wavelength the size of the Milky Way galaxy. This introduces a pressure due to quantum uncertainty and leads to structure suppression on small scales. The case where particles of this nature compose the entirety of the dark matter, which is also referred to as the fuzzy dark matter scenario (FDM), has been thoroughly investigated with $N$-body and hydrodynamical simulations (\citealt{Schive2014CosmicWave,Schwabe2016SimulationsCosmologies,Veltmaat2016CosmologicalMatter,Mocz2017GalaxyHaloes,Veltmaat2018FormationHalos,Nori2018AX-GADGET:Models}). On the other hand, the case where the dark matter is composed of a combination of ULAs and regular cold dark matter (CDM) particles (which we will refer to here as the mixed dark matter case) has been studied only on very large scales via the effects on the matter power spectrum and cosmic microwave background (CMB) anisotropies (\citealt{Amendola2006DarkPseudo-Goldsone-boson,Hlozek2015AData,Kobayashi2017Lyman-alphaUniverse}) and by semi-analytic halo models (\citealt{Marsh2014AMatter}). 
\begin{table*}
  \centering
  \begin{tabular}{|c|c|c|c|c}
    \textbf{Code} & \textbf{Type/Methods} & \textbf{Scales Resolved} & \textbf{Best Uses} & \textbf{Reference}\\
    \hline
    AxionCAMB & Linear PT & Cosmological, linear & Initial density fields/CMB studies & \cite{Hlozek2015AData} \\
    AxionLPT  & Lagrangian PT & Cosmological, linear & Initial position and velocity fields & Present work \\
    AX-GADGET & SPH & Cosmological, greater than $\lambda_\mathrm{dB}$ & Large-scale clustering & \cite{Nori2018AX-GADGET:Models} \\
    AREPO & Pseudospectral & Cosmological $\sim$kpc & Structure formation with baryons & \cite{Mocz2019FirstFilaments,Mocz2019GalaxyGalaxies} \\
    Axion-SPH & SPH & Cosmological $\sim$kpc & Small-scale clustering at $z\gtrsim4$ & \cite{Mocz2015NumericalHydrodynamics} \\
    ENZO  & Hybrid & Cosmological $\sim$kpc & Halo scaling relations at $z\gtrsim4$ & \cite{Veltmaat2018FormationHalos} \\
    GAMER & Pseudospectral & Cosmological $\sim$0.1 kpc & FDM density profiles at $z\gtrsim10$& \cite{Schive2014CosmicWave} \\
    NYX   & Finite-difference & Non-cosmological $\sim$0.1 kpc & FDM halo collisions & \cite{Schwabe2016SimulationsCosmologies} \\
    PYUltralight & Pseudospectral & Non-cosmological $\sim$0.1 kpc & FDM halo collisions & \cite{Edwards2018PyUltraLight:Dynamics} \\
    \hline
  \end{tabular}
  \caption{Overview of existing FDM/ULA algorithms with their most suitable applications ordered from top to bottom by approximate resolution.}
  \label{tab:lit-review}
\end{table*}
A summary of computational techniques used in modeling FDM and ULAs is presented in Table~\ref{tab:lit-review}. 

The standard axion solution to the charge-parity problem of the strong nuclear force predicts just a single axion goldstone boson (\citealt{Peccei1977ConstraintsPseudoparticles,Weinberg1978ABoson,Wilczek1978ProblemInstantons}). Thus the detection of many light axion species would provide support for the existence of compactified extra dimensions and would be a direct probe into the topology of the compact manifold (\citealt{Arvanitaki2010StringAxiverse,Demirtas2018TheAxiverse}). The axion mass distribution has logarithmic support and covers many orders of magnitude (\citealt{Stott2017TheSector}), having the possibility that some axions are light enough to be testable with astrophysical observations. These astronomical constraints complement direct experimental searches for axions or axion-like particles
such as CASPER-ZULF (\citealt{Garcon2019ConstraintsResonance}), nEDM (\citealt{Abel2017SearchFields}), ABRACADABRA (\citealt{Ouellet2019FirstMatter}), ADMX (\citealt{Boutan2018PiezoelectricallyMatter}), CAPP (\citealt{Lee2019CAPP-8TB:mueV}), and QUAX (\citealt{Barbieri2017SearchingProposal}) which probe axions and their coupling to the Standard model in the mass range  $10^{-24}\;\mathrm{eV}\lesssim m \lesssim 10^{-4}\;\mathrm{eV}$. The computational approach presented here contributes to efforts using astrophysical data to establish a lower bound on the mass of the dominant component of the dark matter and probe the possible existence of sub-components.

We investigate the impact of a mixed dark matter model on the large scale dynamics of a type of cosmological simulations based on the Lagrangian perturbation theory (LPT). More specifically, we focus on the introduction of a non-zero sound speed in the equations describing the evolution of individual trajectories in a gravitational potential and in an expanding universe. We build on the work of \cite{Tatekawa2002PerturbationDispersion} who conducted a similar study for matter following a polytropic equation of state. This will require a bit more work as the scale-dependence of the FDM effects are not globally described by a polytrope since they are inherently non-local (they depend on higher order derivatives of the density, not just the value of the density itself). A preview visualization of the results of such a procedure in comparison to the ordinary LPT is presented in Fig.~\ref{fig:mirror}.

First, in Section~\ref{sec:FDM-growth}, we describe the evolution of perturbations for fuzzy and mixed DM scenarios. We also provide an analytic description of the growth factor as a function of axion parameters. Then, in Section~\ref{sec:LPT}, we lay the basis of Lagrangian perturbation theory and introduce the modified growth and quantum pressure in the Zeldovich approximation. In Section~\ref{sec:apps}, we use the modified scheme we created to generate a set of mock initial conditions to evaluate the importance of the quantum pressure at early times for mixed DM models. We additionally examine large scale clustering and shell-crossing to study the impact of the quantum pressure. We also test the hypothesis that the shell-crossing time corresponds the formation of interference fringes by solving the full Schr\"odinger-Poisson system. Finally, we discuss the results in Section~\ref{sec:conclusion}.

%%%%%%%%%%%%%%%%%%%%%%%%%%%%%%%%%%%
%%%%%%%%%%%%%%%%%%%%%%%%%%%%%%%%%%%

\section{Linear Growth Factor for Fuzzy Dark Matter}\label{sec:FDM-growth}

\subsection{Dark Matter as a Light Scalar Field}
ULAs are described by a non-relativistic scalar field, $\varphi$ with action (\citealt{Hui2017UltralightMatter})
\begin{align}
    S_\varphi = \int\frac{d^4x}{\hbar} \sqrt{-g} \bigg[\frac{1}{2}g^{\mu\nu}\partial_\mu \varphi \partial_\nu \varphi -  V(\varphi) \bigg],
\end{align}
where $g^{\mu\nu}$ is the FLRW metric, $g$ is its determinant, and $V$ is the field potential. As it is often the case when discussing light dark matter particles, we pick units where $c=1$, but where the $\hbar$ dependency is kept explicit. Considering a model without self-interactions, we have a potential of the form
\begin{align}
    V(\varphi) = \frac{1}{2}\frac{m^2}{\hbar^2}\varphi^2. 
\end{align}
The field then obeys the Klein-Gordon equation coupled to the Einstein field equations (\citealt{Widrow1993UsingMatter}). Under the assumption that the field is in the non-relativistic regime, we arrive at an axion wavefunction $\psi$ as
\begin{align}
    \varphi = \sqrt{\frac{\hbar^3}{2m}} \Big(\psi e^{-imt/\hbar} + \psi^* e^{imt/\hbar}\Big).
\end{align}
Following the de Broglie-Bohm approach, we first write the Schr\"odinger equation for the wavefunction which gives
\begin{align}
    i\hbar \bigg(\frac{\pa\psi}{\pa t} +\frac{3}{2}H\psi\bigg)= -\frac{\hbar^2}{2ma^2}\nabla^2 \psi + m\Phi\psi, \label{eq:Schro}
\end{align}
where $H=\dot{a}/a$ is the Hubble rate and where $\Phi$ is the gravitational potential following Poisson's equation. This equation is coupled with gravity through its potential obeying Poisson's equation
\begin{align}
    \nabla^2 \Phi = 4\pi G \rho. \label{eq:Poisson}
\end{align}
Then we write (using the Madelung form of \citealt{Madelung1926EineSchrodinger}) that
\begin{align}
    \psi = |\psi|e^{i\theta},
\end{align}
where $|\psi|^2 =  \frac{\rho}{m}$ and $\mathbf{v} = \frac{\hbar}{ma}\nabla \theta$. Using the latter in the Schr\"odinger equation, and taking the real and imaginary parts, one arrives at fluid equations
\begin{align}
    \frac{\pa \rho}{\pa t} + 3H\rho + \frac{1}{a}\nabla(\rho\mathbf{v}) &= 0,\\
    \frac{\pa \mathbf{v}}{\pa t} + H \mathbf{v} + \frac{1}{a} \mathbf{v}\cdot\nabla\mathbf{v} &= -\frac{1}{a}\nabla(\Phi+Q), \label{eq:Euler}
\end{align}
where we define the quantum pressure\footnote{It is worth noting that the widely used term of ``quantum pressure" is not quite accurate as $Q$ is a potential and as the system behaves following the equation of motion of a classical field given the very high particle occupation number.}
\begin{align}
    Q \equiv -\frac{\hbar^2}{2m^2a^2}\frac{\nabla^2 \sqrt{\rho}}{\sqrt{\rho}} \label{eq:QP}.
\end{align}

%%%%%%%%%%%%%%%%%%%%%%%%%%%%%%%%%%%

\subsection{Linear Growth with a Sound Speed}
To see how the clustering of FDM particles differs from the CDM case, we first examine some elements of Eulerian (cosmological) perturbation theory in the non-relativistic regime. The main object of study is the overdensity denoted 
\begin{align}
\delta_I(\x,t)\equiv \rho_I(\x,t)/\bar{\rho}_I(t) -1,
\end{align}
where $\bar{\rho}_I(t) = \bar{\rho}_{I,0} a(t)^{-3}$ is the background density and where $I$ is a subscript for the particle species (ultralight axions, baryons, CDM, etc.). In Eulerian perturbation theory, the underlying assumption is that the overdensities and the velocities are small: $\delta,\: |\mathbf{v}|\ll 1$, and that any second order term involving either can be neglected. In this regime, the fluid equations for the overdensity take the form (for Eulerian perturbation theory for CDM see \citealt{Peebles1980Large-ScaleUniverse,Padmanabhan1993StructureUniverse,Bouchet1995IntroductoryTheories} and references therein, while more details on Eulerian perturbation of FDM applied to the matter power spectrum see \citealt{Li2019NumericalModel})
\begin{align}
    &\dot{\delta}_I + a \nabla \cdot \mathbf{v}_I = 0,
    \\&\ddot{\delta}_I + 2\frac{\dot{a}}{a}\dot{\delta}_I   =\frac{\nabla^2 P_I}{a^2 \bar{\rho}_I} + 4\pi G \bar{\rho}_m \sum_I \frac{\Omega_I}{\Omega_m} \delta_I, \label{eq:oddP}
\end{align}
where $P$ corresponds to the pressure of the matter fluid. From here on, we consider only ultralight axions (FDM) and CDM and leave the treatment of baryons to future work. Therefore, we will work under the condition that $\Omega_m =\Omega_\mathrm{DM}=\Omega_a + \Omega_\mathrm{CDM}$, where $\Omega_I$ is the mean density in species $I$ over the critical density. This means that Eq.~(\ref{eq:oddP}) describes a set of two coupled differential equations one of which (the CDM equation) has vanishing pressure. Generally, Eq.~(\ref{eq:oddP}) is expressed in a more familiar form assuming a relation $P=P(\rho)$, in which case $P \approx P(\bar{\rho}) + \frac{dP}{d\rho} \bar{\rho} \delta $. Under this description, we obtain for axions
\begin{align}
  \ddot{\delta}_a + 2\frac{\dot{a}}{a}\dot{\delta}_a   =\frac{c_s^2 \nabla^2\delta_a}{a^2} + 4\pi G \bar{\rho}_m \delta_\mathrm{DM}, \label{eq:P&Sound}
\end{align}
where the sound speed is defined as $c_s^2 = \frac{dP}{d\rho}$ and where
\begin{align}
    \delta_\mathrm{DM} \equiv \frac{\Omega_a}{\Omega_\mathrm{DM}} \delta_a + \bigg(1-\frac{\Omega_a}{\Omega_\mathrm{DM}} \bigg)\delta_\mathrm{CDM}.
\end{align}
In the case of ULAs, there is a pressure term which leads to a non-vanishing sound speed given customarily in Fourier space by (\citealt{Marsh2016AxionCosmology})
 \begin{align}
    c_s^2 = \frac{\hbar^2k^2}{4m^2a^2},
    \label{eq:soundspeed}
 \end{align} 
where $k$ is the comoving wavenumber. When considering pressureless matter (also referred to as dust) in an Einstein-de Sitter Universe, Eq.~(\ref{eq:oddP}) can be solved with the ansatz
\begin{align}
    \delta(\x,a) = C_+(\x)D_+(a) + C_-(\x)D_-(a),
\end{align}
where $D_+$ and $D_-$ are respectively referred to as the growing and decaying modes of the linear growth factor. In this case, the equation has an analytical solution for the linear growth factor which corresponds to $D_+(a)\propto a$ and $D_-(a)\propto a^{-3/2}$. We will come back to this scenario when considering the impact of the axion pressure on the linear growth in the Eulerian case.
In the fully FDM case ($\Omega_a=\Omega_\mathrm{DM}$), we find the axion linear growth factor is scale dependent and its evolution follows
\begin{align}
    \ddot{D}(k,a) + 2H\dot{D}(k,a) + \bigg(\frac{\hbar^2k^4}{4m^2a^4} - 4\pi G \bar{\rho}_a \bigg)D(k,a) = 0. \label{eq:PDE_growth}
\end{align}
The scale for which the two terms in the brackets in the third term are equal is known as the axion Jeans scale and is given by (\citealt{Hu2000FuzzyParticles,Marsh2016AxionCosmology})
\begin{align}
    k_J = 66.5a^{1/4}\bigg(\frac{m}{10^{-22}\;\mathrm{eV}}\bigg)^{1/2} \bigg(\frac{\Omega_\mathrm{DM} h^2}{0.12}\bigg)^{1/4}\;\mathrm{Mpc}^{-1}. \label{eq:kJeans}
\end{align}
We have an analytic solution for the growing mode of Eq.~(\ref{eq:PDE_growth}) in the case $k\gg k_J$ yielding (\citealt{Marsh2016AxionCosmology})
\begin{align}
    D_+(k,a) = &\frac{3mH_0\sqrt{a}}{\hbar k^2}\sin\bigg(\frac{\hbar k^2}{mH_0\sqrt{a}}\bigg)\nonumber \\&+ \bigg(\frac{3m^2H_0^2 a}{\hbar^2k^4}-1\bigg)\cos\bigg(\frac{\hbar k^2}{mH_0\sqrt{a}}\bigg). \label{eq:marsh_grow}
\end{align}

The study of the above solution allows us to determine the ranges of particle masses, scales, and redshifts for which the modified scale factor is indistinguishable from the usual Eulerian case.
\begin{figure}
    \centering
    \includegraphics[width=\linewidth]{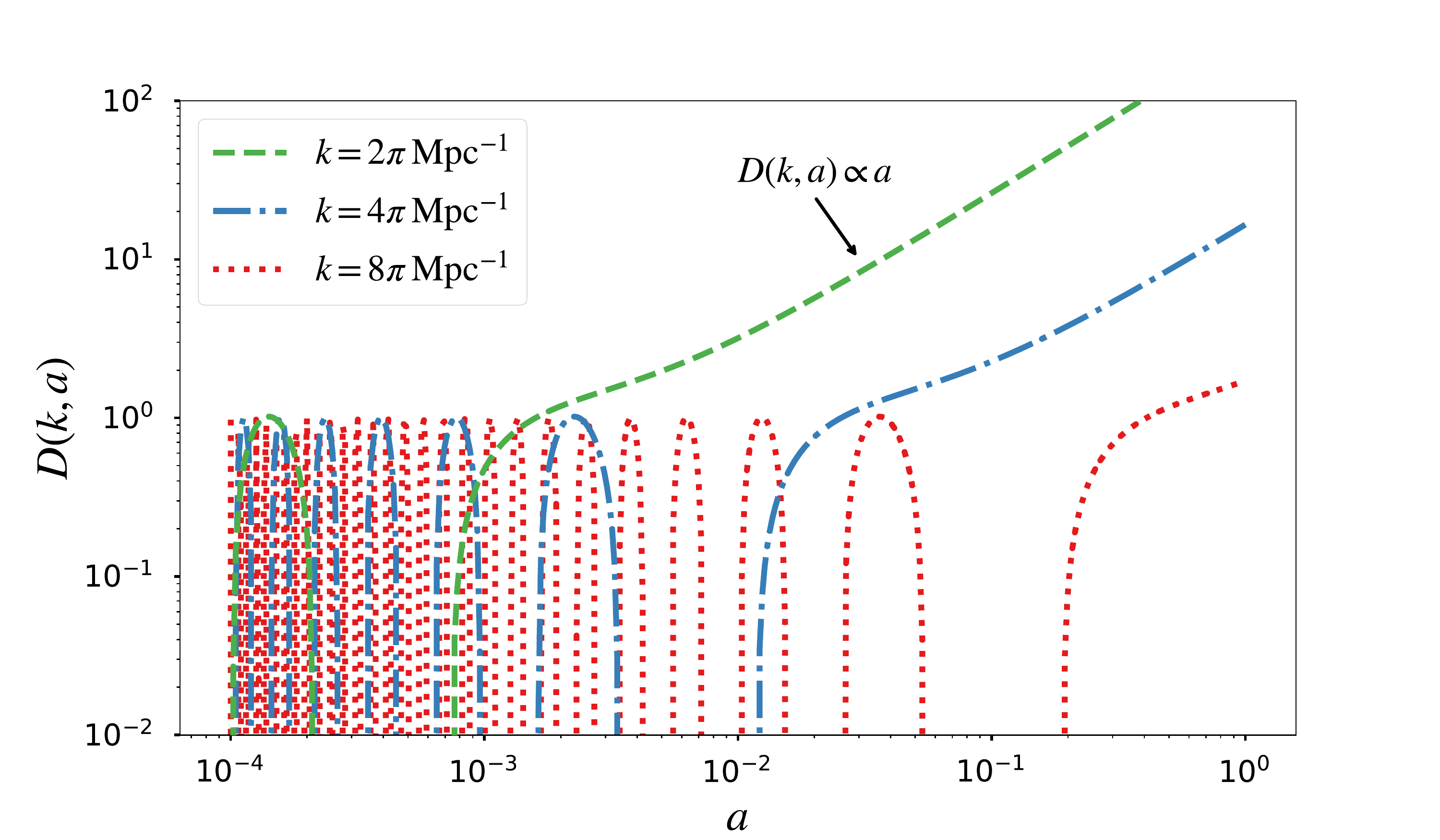}
    \caption{Modified linear growth factor as a function of the scale factor at different length scales for an axion mass of $m= 10^{-23}$ eV. The axions grow like standard CDM at late times $z<10, a>0.1$ on $k=4\pi$Mpc$^{-1}$ scales (blue dashed lines), while on smaller scales the axions exhibit oscillatory behaviour for all redshifts (red dotted lines).}
    \label{fig:LGF}
\end{figure}
The results are plotted for different scales in Fig. \ref{fig:LGF} where we observe the same oscillatory behaviour of the linear growth factor at high redshift that was noticed in \cite{Hlozek2015AData}. From these plots we can estimate, for instance, that for an axion mass of $m=10^{-23}$ eV and a scale $k=4\pi\:\mathrm{Mpc}^{-1}$, the evolution of the FDM density perturbations have the same growing mode as the CDM case for redshifts $z<10$. \cite{Li2019NumericalModel} also finds an expression for the growth as a ratio of Bessel functions of fractional order, $J_{-5/2}$, in a matter dominated universe which takes the form
\begin{align}
    D(k,a) = \bigg(\frac{a_i}{a}\bigg)^{1/4}\frac{J_{-5/2}\Big(\hbar k^2/mH_0\sqrt{a}\Big)}{J_{-5/2}\Big(\hbar k^2/mH_0\sqrt{a_i}\Big)}\label{eq:li_grow},
\end{align}
where $a_i$ is the scale factor corresponding to a high initial redshift and where
\begin{align}
    J_{-5/2}(x) =  \sqrt{\frac{2}{\pi x}}\bigg(\frac{3\cos x}{x^2}+\frac{3\sin x}{x}-\cos x\bigg).
\end{align}
This expression shows the same behaviour of growth on large scales and fast oscillations on small scales as that of Eq. (\ref{eq:marsh_grow}).
\begin{figure}
    \centering
    \includegraphics[width=1.00\linewidth]{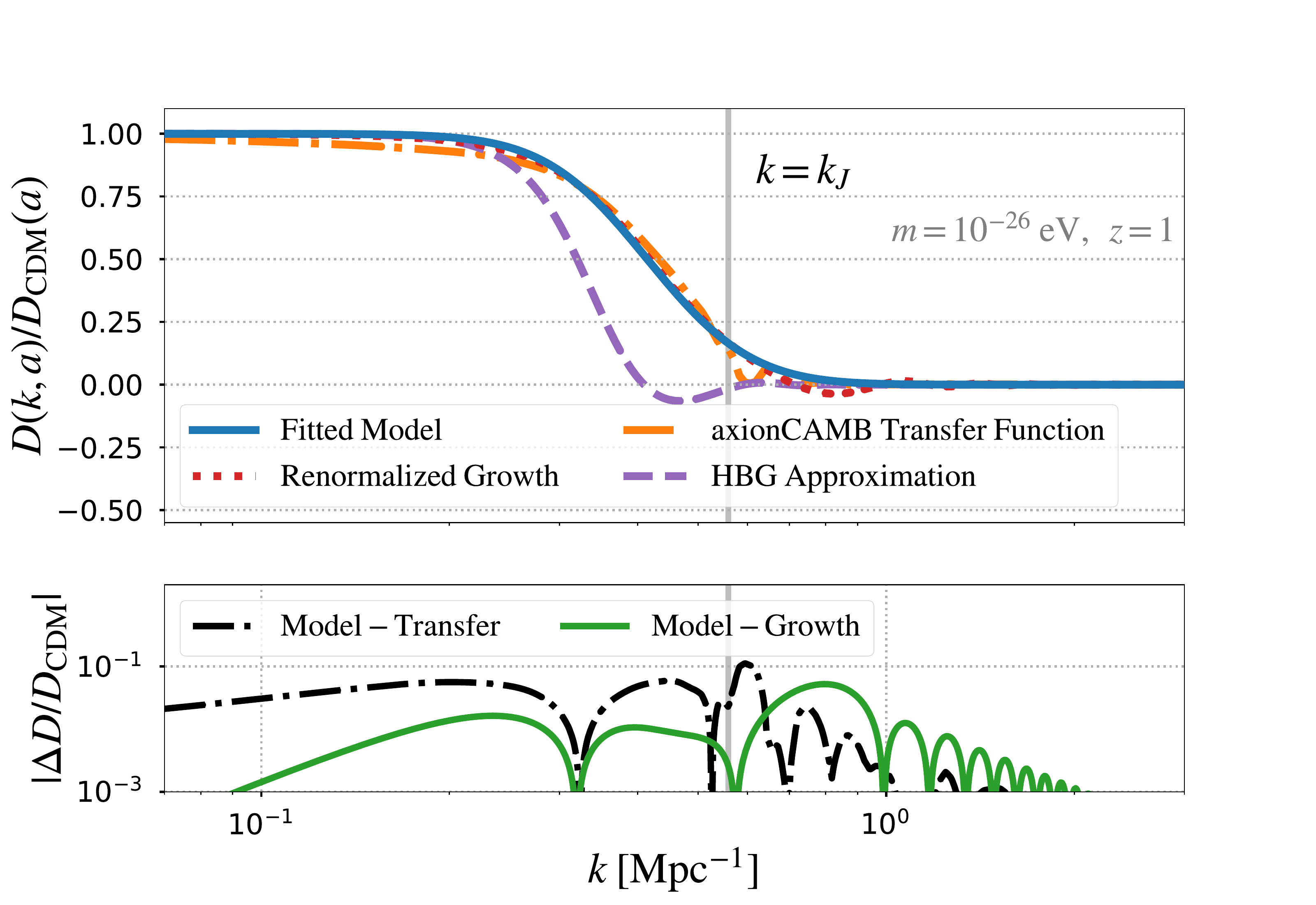}
    \caption{Linear growth factor in Fourier space in the presence of a scale-dependent sound speed. On scales smaller than the Jeans scale the axions oscillate around the zero point. The fitted model is obtained by fitting the renormalized growth with Eq.~(\ref{eq:model}). The renormalized growth is taken by expanding the denominator of Eq.~(\ref{eq:li_grow}) with Eq.~(\ref{eq:taylor_exp}). The HBG approximation is taken from Eq.~(\ref{eq:HBG}) while the axion transfer function Eq.~(\ref{eq:transfer_f}) is obtained from \texttt{axionCAMB}. We observe that the renormalized growth, transfer function and smoothed Heaviside methods of solving Eq.~(\ref{eq:PDE_growth}) are in good agreement from the residuals plot (bottom panel). %\renee{Need to clear}
    }
    \label{fig:growth_fit}
\end{figure} We note however, that the solutions obtained in Eq.~(\ref{eq:marsh_grow})-(\ref{eq:li_grow}) are rapidly oscillating on small-scales. For our purposes, we wish to obtain an \emph{averaged} growth for ease of computation and set the growth at large wavenumbers to zero. The model we use for the mean growth takes the form of a smoothed Heaviside step function with shape
\begin{align}
    D(k,a) &\approx \bigg(1 -\Big 
    [1+e^{-2\alpha(k-k_0)}\Big]^{-8} \bigg) D_\mathrm{CDM}(a) \label{eq:model}
    \\&\equiv L(k)D_\mathrm{CDM}(a), \label{eq:lp}
\end{align}
where $D_\mathrm{CDM}$ is the growth factor in the pure CDM case solving Eq. (\ref{eq:PDE_growth}) when taking $c_s^2=0$.

In order to calculate the values of the parameters $\alpha$ and $k_0$, we must first find the appropriate value of $a_i$. As shown in Fig.~\ref{fig:growth_fit}, the transition between the decaying and oscillating regimes of the linear growth factor is the singular point where $J_{-5/2}\Big(\hbar k^2/mH_0\sqrt{a_i}\Big) = 0$\footnote{This does not cause a divergence in the plotted growth factor thanks to an alternate choice of normalization presented in more details in Eq.~(\ref{eq:taylor_exp}), but the singular point is nevertheless present in the full form of Eq.~(\ref{eq:li_grow}).}. Mathematically, the location of this point is governed by the value of the initial redshift for the integration, but physically it corresponds to the Jeans scale of the particles of Eq.~(\ref{eq:kJeans}). Thus, we can impose $k=k_J$ at this point and solve numerically for the physically appropriate value of the free parameter $a_i$. After this, we normalize the resulting $D(k,a)$ by the CDM solution $D_\mathrm{CDM}(a)$. To verify the assumption that the Jeans scale does correspond to the first oscillation of the denominator of Eq.~(\ref{eq:li_grow}), we generate the matter power spectrum at redshift $z=1$ for a universe with DM made of pure FDM at mass $m=10^{-26}$ eV to match the case portrayed in Fig.~\ref{fig:growth_fit}. For this, we use the publicly available Boltzmann solver \texttt{axionCAMB}~(\citealt{Lewis2000CAMB,Hlozek2015AData}). The code offers a solution to Eq.~(\ref{eq:PDE_growth}) at matter radiation equality so in order to obtain the $z=1$ equivalent solution, we employ the scaling relation of the Jeans scale $k_J(a) \propto a^{1/4}$ which we use as a translation given by
\begin{align}
    k\to\tilde{k} \equiv \bigg(\frac{a_\mathrm{eq}}{a}\bigg)^{1/4} k, \label{eq:scaling_z}
\end{align}
where $a_\mathrm{eq}$ is the scale factor at matter-radiation equality. In order to make an appropriate comparison, we normalize the result by the CDM power spectrum and obtain the FDM transfer function ($T^2_\mathrm{F}$) at $z=1$ given by
\begin{align}
    T^2_\mathrm{F}\big(\tilde{k}\big) = \frac{P_\mathrm{FDM}\big(\tilde{k}\big)}{P_\mathrm{CDM}\big(\tilde{k}\big)}.\label{eq:transfer_f}
\end{align}
We also use a semi-analytic expression for the FDM transfer function from \cite{Hu2000FuzzyParticles} which we will refer to as the ``HBG approximation" which reads
\begin{align}
    T_\mathrm{F}\big(\tilde{k}\big) \approx \frac{\cos\big[\big(A\tilde{k}\big)^3\big]}{1+\big(A\tilde{k}\big)^8}, \label{eq:HBG}
\end{align}
where $A=0.179(m/{10^{-22}\;\mathrm{eV}})^{-4/9}\;\mathrm{Mpc}$. Both the translated transfer functions and the HBG approximations are plotted in Fig.~\ref{fig:growth_fit} where we can see a good agreement between the growth factor fitting technique of the current work and the \texttt{axionCAMB} numerical result.

We note two major differences between the transfer function approximation and the expression of Eq.~(\ref{eq:li_grow}) which are the absence of singularities and oscillations. The first are unphysical and this can be quickly seen by comparing the solutions of Eq.~(\ref{eq:marsh_grow}) and (\ref{eq:li_grow}) at large $k$. One exhibit bounded oscillations while the other shows a divergence when the denominator (which is also oscillatory) hits zero. This is simply a matter of choice of normalization factor which can be rescaled or smoothed over. The physical boundary conditions $\lim_{k\to0} D(k,a)/D_\mathrm{CDM} =1$ and $\lim_{k\to\infty} D(k,a)/D_\mathrm{CDM} =0$ coupled with our understanding of the Jeans scale enable us to simply remove the singular points from the growth factor. An alternative normalization which gets rid of divergences is found by Taylor expansion of the denominator to the first three positive terms (any other number of terms will allow fast oscillations and/or divergences). In other words, the denominator of Eq.~(\ref{eq:li_grow}) can be approximated by
\begin{align}
    J_{-n}(x) = x^{-n}\bigg[&\frac{2^n}{\Gamma(1-n)} + \frac{2^{n-2}x^2}{(n-1)\Gamma(1-n)}  \nonumber \\&+\frac{2^{n-5}x^4}{(n-2)(n-1)\Gamma(1-n)} 
    + \mathcal{O}(x^6)\bigg], \label{eq:taylor_exp}
\end{align}
where $\Gamma(x) \equiv \int_0^\infty dy y^{x-1}e^{-y}$ is the Euler Gamma function. Normalizing with the above instead gives little oscillations and no divergences. Then, we also smooth over the oscillations of the numerator which are in this case physical. These oscillations also appear when studying a fixed-scale time-varying solution such as in Fig.~\ref{fig:LGF}. It is physically valid to neglect them as an oscillating modes below the axion Jeans scale are no longer growing. This is similar to the oscillations seen with baryons with the exception that here the pressure does not follow a polytropic equation of state. The quantum pressure is non-local and has a greater scale dependence. We study the impact of these oscillations on the particles' LPT displacements in Appendix~\ref{sec:app_osc}.

Now that we've established a good correspondence between the transfer function and the solution for the axion growth, we obtain semi-analytic expressions for the free parameters $\alpha$ and $k_0$. We fit the \texttt{axionCAMB} transfer function at various ratios of $\Omega_a/\Omega_\mathrm{DM}$ with the $L(k)$ model described in Eq.~(\ref{eq:model}). We express the characteristic scale as a function of the Jeans scale of Eq.~(\ref{eq:kJeans}) as
\begin{align}
    k_0 \approx 0.0334\bigg(\frac{m}{10^{-24}\;\mathrm{eV}}\bigg)^{-0.00485} \bigg(\frac{\Omega_a}{\Omega_\mathrm{DM}}\bigg)^{0.527} k_J. \label{eq:k0_scale}
\end{align}
We also fit for the mass and fraction dependence of $\alpha$ giving
\begin{align}
    \alpha \approx 0.194\bigg(\frac{m}{10^{-24}\;\mathrm{eV}}\bigg)^{-0.501} \bigg(\frac{\Omega_a}{\Omega_\mathrm{DM}}\bigg)^{0.0829} \; \mathrm{Mpc}. \label{eq:alpha_scale}
\end{align}
Note that we express the scaling with respect to $m/10^{-24}\;\mathrm{eV}$ here as our fitting procedure was made for the range $10^{-27} \;\mathrm{eV}\leq m \leq 10^{-24} \;\mathrm{eV}$. We assumed no redshift dependence on the slope $\alpha$. This assumption is tested by translating wavenumbers using Eq.~(\ref{eq:scaling_z}) and comparing the linear transfer function slope to the solution of Eq.~(\ref{eq:li_grow}). We found good agreement between the two as shown in Fig.~\ref{fig:growth_fit} and deduce that the change in slope from $z=z_\mathrm{eq}$ to $z=1$ is negligible. From our search in linear theory, we have seen that the linear growth factor for FDM differs on small scales from that of CDM due to the presence of a quantum pressure term. Solving the full Eq.~(\ref{eq:PDE_growth}) numerically leads to rapidly oscillating functions and is potentially unstable. Hence, we developed a numerical procedure to obtain a smooth curve quantifying the evolution of FDM linear growth factor as a function of the cosmological and axion parameters $\{a,\;m,\;\Omega_a\}$, allowing us to consider mixed DM scenarios with ease.

%%%%%%%%%%%%%%%%%%%%%%%%%%%%%%%%%%%
%%%%%%%%%%%%%%%%%%%%%%%%%%%%%%%%%%%

\section{Beyond Linear Theory} \label{sec:LPT}
In the previous section, we have described the linear perturbation theory results describing the growth of structure in CDM, FDM and mixed DM scenarios. To go beyond this usual treatment, we change from Eulerian to Lagrangian coordinates. The Lagrangian perturbation theory approach tracks the trajectories of individual test particles rather than evolution of the density field. The main object of study is the displacement vector, $\p$ which relates the final (Eulerian) positions $\x$ to the initial (Lagrangian) positions $\q$ following
\begin{align}
\x(\q,\tau) = \q + \p (\q,\tau),
\end{align}
where we use the conformal time $dt = a d\tau$. By mass conservation, we have that the determinant of the Jacobian matrix $J$ of coordinate change follows \cite{Jeong2010CosmologySurveys}
\begin{align}
    \bar{\rho} d^3q = \rho(\x)d^3x = \bar{\rho}(1+\delta(\x))d^3x,
\end{align}
and therefore
\begin{align}
    1 + \delta(\x) = \bigg|\frac{d^3q}{d^3x} \bigg| = \frac{1}{J(\q)}. \label{eq:jacobian}
\end{align}
This allows the definition of the Jacobian matrix determinant as a function of the displacement
\begin{align}
    J(\q) \equiv \det\bigg[\delta_{ij}+\frac{\partial \Psi_i}{\partial q_j}\bigg],
\end{align}
where $\delta_{ij}$ is the Kronecker delta. The particles' path as a function of time in the pressureless CDM case follows
\begin{align}
    \frac{d^2 \x_\mathrm{CDM}}{d\tau^2}+2\mathcal{H} \frac{d \x_\mathrm{CDM}}{d\tau}  = -\nabla_\x \Phi,
\end{align}
and for the axions with a non-zero sound speed (\citealt{Tatekawa2002PerturbationDispersion})
\begin{align}
    \frac{d^2 \x_a}{d\tau^2}+2\mathcal{H} \frac{d \x_a}{d\tau} = -\nabla_\x \Phi + \frac{c_s^2}{a^2}\nabla_\x\delta_a,
\end{align}
where $\mathcal{H}\equiv aH$. By taking the gradient of both sides, we can use Poisson's equation and Friedmann's equation
\begin{align}
    H^2 = \frac{8\pi G}{3}\rho,
\end{align}
where $\rho = \sum_I \rho_I$ which gives
\begin{align}
   \nabla_\x \cdot \bigg( \frac{d^2 \x_\mathrm{CDM}}{d\tau^2} &+ 2\mathcal{H} \frac{d \x_\mathrm{CDM}}{d\tau}\bigg) = -\frac{3}{2}\mathcal{H}^2 \Omega_m \delta_\mathrm{DM} , \label{eq:traj_cdm}
\end{align}
and
\begin{align}
   \nabla_\x \cdot \bigg( \frac{d^2 \x_a}{d\tau^2} &+ 2\mathcal{H} \frac{d \x_a}{d\tau}\bigg) = -\frac{3}{2}\mathcal{H}^2 \Omega_m \delta_\mathrm{DM} + \frac{c_s^2}{a^2}\nabla^2_\x \delta_a . \label{eq:traj}
\end{align}
From the relation between the effective sound speed and the equation of state $c^2_s=\frac{d P}{d\rho}$, we will obtain the form of the displacement $\p$ in Lagrangian space with a non-zero pressure.
We consider the Zeldovich approximation (\citealt{ZelDovich1970GravitationalPerturbations.}) which is the first order solution to the equation of motion
\begin{align}
    \p = \p^{(1)} + ...
\end{align}
In this approximation, we can write the determinant of the Jacobian matrix as
\begin{align}
    \frac{1}{J(\q)}= 1 - \nabla_\q \cdot \p^{(1)}. \label{eq:disp_poisson}
\end{align}
Assuming an irrotational displacement field ($\nabla_\q \times \p^{(1)} = 0$), we have that the displacement can be expressed as the gradient of the \textit{displacement potential} which we denote by $\phi^{(1)}$. By combining this expression with Eq.~(\ref{eq:jacobian}) and Eq.~(\ref{eq:disp_poisson}), we obtain the Poisson equation
\begin{align}
    \p^{(1)} = \nabla_\q \phi^{(1)}\Rightarrow \nabla^2_\q \phi^{(1)} = \delta(\q). \label{eq:scalar_poisson}
\end{align}
In the pressureless case, it is customary to multiply the solution $\phi^{(1)}$ directly with the linear growth factor $D_\mathrm{CDM}$(a). In the case of FDM, however, the linear growth is scale-dependent and the prescription used obtain a full solution is somewhat different. As Eq. (\ref{eq:scalar_poisson}) is solved with spectral methods, we multiply the solution before taking the inverse Fourier transform back to real space. In other words the modified solution is found by
\begin{align}
    \tilde{\p}^{(1)}(\q,a) =  \mathcal{F}^{-1}\bigg\{ \frac{-i\mathbf{k} L(k)}{k^2}\mathcal{F}\{\delta(\q) \} \bigg\} D_\mathrm{CDM}(a),\;\label{eq:modified_disp}
\end{align}
where $\mathcal{F}$ denotes the Fourier transform and $\tilde{\p}$ is the modified displacement. $L(k)$ is the smoothed Heaviside approximation to the modified axion growth factor. The semi-analytic expression for $L(k)$ was shown in Eq.~(\ref{eq:lp}). The scheme of Eq.~(\ref{eq:modified_disp}) can be extended beyond the Zeldovich approximation to second order LPT as well. In this case, one would simply calculate the first and second order displacements and apply the low-pass filter to each so that
\begin{align}
    \tilde{\p}^{(2)}(\q,a) = \mathcal{F}^{-1}\{L(k)\} \star \p^{(2)}(\q,a).
\end{align}
The justification for this extension is given in detail in Appendix~\ref{sec:appendix-2lpt}.

One must be careful when taking the Fourier transform of a function in Lagrangian coordinates as the wavenumbers have a different physical meaning than in Eulerian space. To verify that the expression in Eq.~(\ref{eq:soundspeed}) is also valid in Lagrangian space. The Eulerian and Lagrangian wavenumbers are equivalent to the Laplacian operator with respect to $\q$ or $\x$ and the change in variables between the two coordinate systems is done through the deformation tensor
\begin{align}
    \mathcal{D}_{ij} &\equiv \frac{\partial x_i}{\partial q_j} =\delta_{ij} + \frac{\partial \Psi_i}{\partial q_j}, \label{eq:deform_tens}
\end{align}
where $\delta_{ij}$ is the Kronecker delta. Assuming the spatial variation is coarse grained relative to the inverse wave number,the deformation (phase conserving) obeys 
\begin{align}
    k_\mathrm{Lagrange} dq^i &= k_\mathrm{Euler} dx^i
    \\&= k_\mathrm{Euler}\mathcal{D}_{ij} dq^j
    \\&= k_\mathrm{Euler} \bigg(dq^i + \frac{\partial \Psi_i}{\partial q_j} dq^j\bigg).
\end{align}
However, by Eq.~(\ref{eq:scalar_poisson}), we have that $ \partial_q \Psi^{(1)}  = \mathcal{O}(\delta)$, therefore we have at first order
\begin{align}
    k_\mathrm{Lagrange} dq^i = k_\mathrm{Euler} dq^i + \mathcal{O}(\delta),
\end{align}
and so we expect the equivalence between Eulerian and Lagrangian wavenumbers to hold for small density perturbations (as is the case for high redshift initial conditions).

%%%%%%%%%%%%%%%%%%%%%%%%%%%%%%%%%%%
%%%%%%%%%%%%%%%%%%%%%%%%%%%%%%%%%%%

\section{Applications} \label{sec:apps}
Constraints on FDM found in the literature often allow for a mixed dark matter composed of a combination of CDM and FDM parameterized by a FDM fraction between 0 and 1. Depending on the mass range of the FDM particle mass and the constraining method used, constraints can be easier or harder to reach. For ``heavy" particles ($10^{-23}-10^{-21}$ eV), the capacity for constraints becomes limited when the FDM fraction goes below 25\%, while for lighter masses down to $10^{-27}$ eV, the constraints tend to be tighter at the few-percent level. This is due to the linear matter power spectrum being only partially suppressed over a long range of scales at low FDM fraction (see for example \citealt{Marsh2013AxiverseInflation}). Here we demonstrate that modifications to LPT are essential to probe these dark matter configurations either by allowing accurate computations for initial conditions or by providing an efficient large-scale cosmic web modeling tool. To generate large-scale structure maps and velocity flows from a choice of parameters \{$a,\;m,\;\Omega_a$\}, we use the following 5 step procedure:
\begin{enumerate}
    \item Generate a power spectrum of density fluctuations using an adapted Boltzmann code (e.g. \texttt{axionCAMB}),
    \item create a realization of a density field based on the power spectrum of step (i),
    \item calculate the CDM displacements using ordinary LPT,
    \item calculate the values of $\alpha$ and $k_0$ from the redshift and axion parameters,
    \item convolve the displacement field of step (iii) with $L(k)$ using the values $\alpha$ and $k_0$ found in step (iv).
\end{enumerate}

 \subsection{Adapted Initial Conditions}
The cosmological simulation codes listed in the five middle rows of Table~\ref{tab:lit-review} need initial conditions generated from LPT. If we wish to use these computational methods to assess the implications of mixed fuzzy dark matter scenarios, it becomes imperative to adapt LPT to generate self-consistent initial conditions for all light dark matter compositions due to the possible formation of transients (\citealt{Scoccimarro1997TransientsAnalysis}). Note that currently all cosmological simulations use the linear power spectrum (sometimes even without FDM effects, if the starting $z$ is high enough), but none use the adapted LPT.
\begin{figure}
    \begin{subfigure}
        \centering
        \includegraphics[trim = 0mm 10mm 0mm 0mm,clip,width=.96\linewidth]{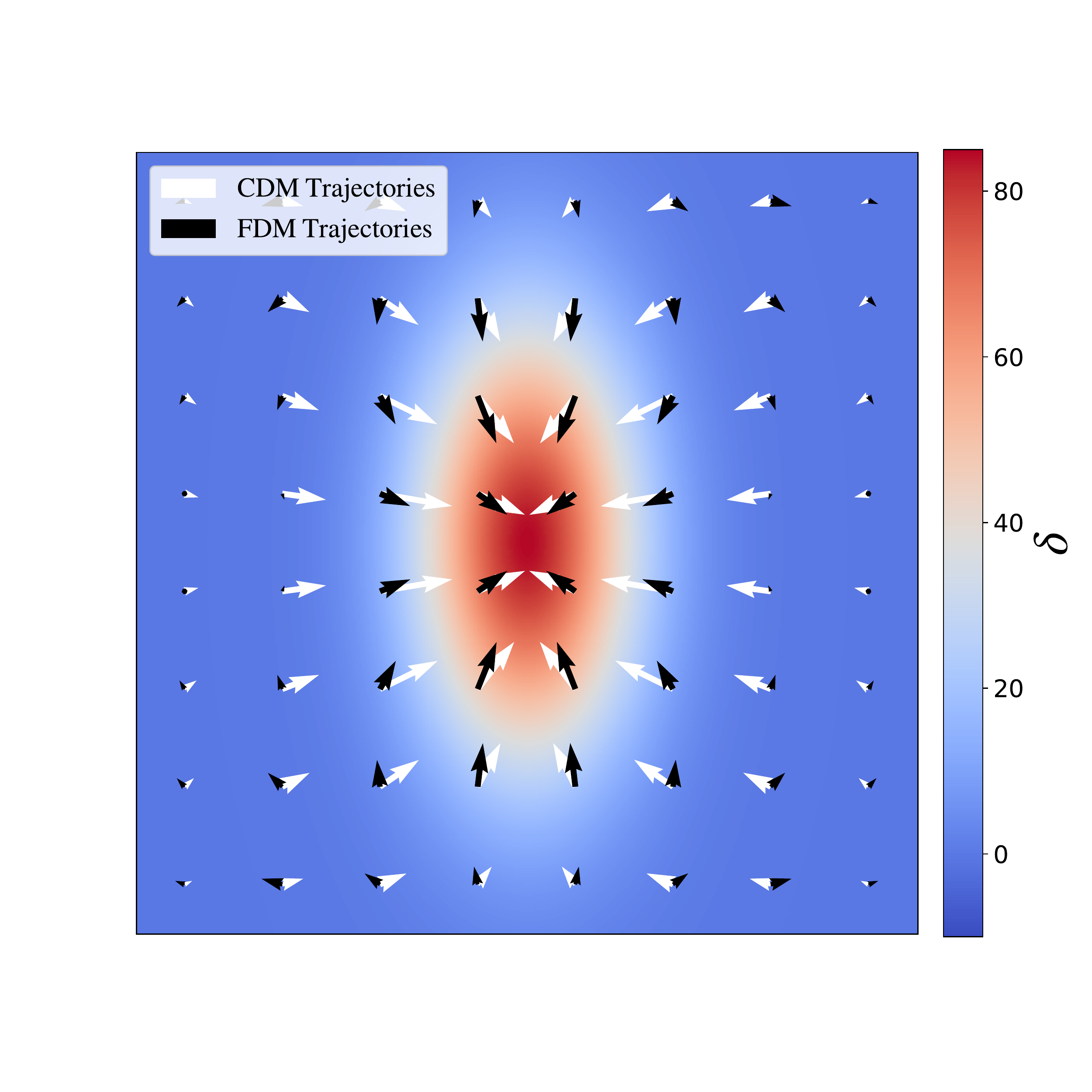}
    \end{subfigure}
    \vskip -30pt%
    \begin{subfigure}
        \centering
        \includegraphics[trim = 0mm 25mm 0mm 18mm, clip, width=.96\linewidth]{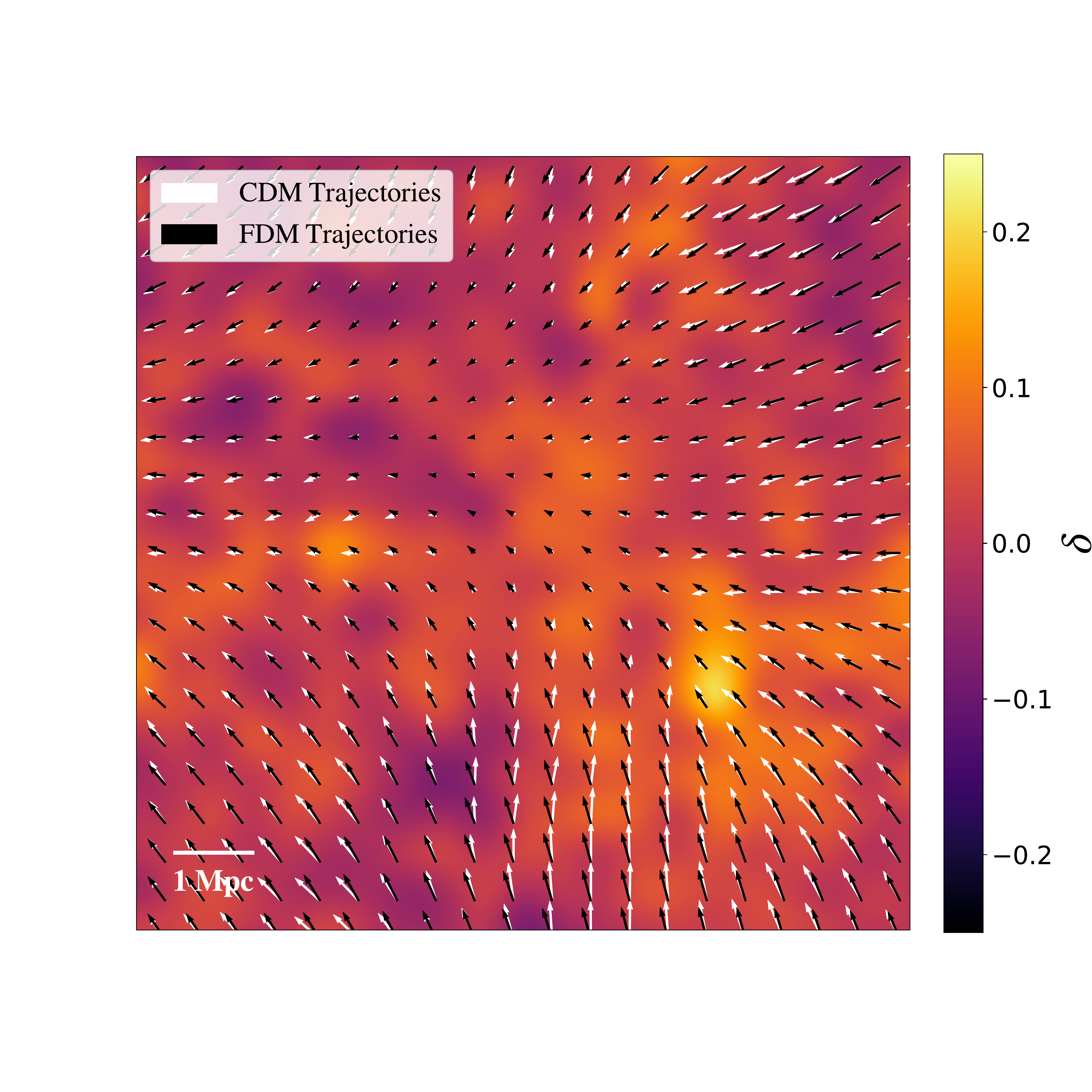}
    \end{subfigure}
    \caption{(\textit{Top}) Toy example of displacements around a 2D Gaussian overdensity with $\sigma_x<\sigma_y$. The magnitude of the FDM displacements is shortened around the regions of high density and the angle of the displacement vectors change in two instances. (\textit{Bottom}) Physical illustration of the displacement field at redshift $z=80$ resulting from the underlying initial density field. The latter was obtained with a modified matter power spectrum with 10\% of $10^{-27}$ eV axions. The displacements for the CDM particles are obtained using the regular LPT scheme while the axions' have been determined using the modified approach of Eq.~(\ref{eq:modified_disp}). All the displacements have been scaled up for ease of visualization.
    }
    \label{fig:disp_z80}
\end{figure}
To test the implications of the scale-dependent pressure on Lagrangian displacements, we create a $512^3$ grid particles subject to the gravitational attraction of an initial density field. The displacement fields of the uniformly distributed test particles for both the CDM and FDM cases are presented in Fig.~\ref{fig:disp_z80}. When comparing the trajectories for the CDM and FDM particles in our toy example, we see that both the amplitude and the direction are affected with the most notable differences occurring around regions of peak density. We note a preferential flow away from overdensities for FDM which is the manifestation of the quantum pressure on the particle trajectories. Conversely, the CDM flows are strictly dictated by the gravitational potential. Furthermore, due to the redshift dependence found in Eq.~(\ref{eq:kJeans}), the effects of the quantum corrections on the LPT increase in importance at high redshift. Most high resolution simulation schemes obtain initial conditions at $z\sim 100$ including an initial velocity field and it is easy to see that some care is necessary to avoid inconsistent initial conditions in the presence of a quantum pressure and multiple particle species. From Fig.~\ref{fig:disp_z80}, we observe that (1) the FDM trajectories are generally shortened in comparison to CDM, and (2) the deflections in the FDM trajectories ``push" them to be slightly oriented away from overdensities in a manner consistent with the results of other FDM simulation techniques.

Setting initial conditions without taking this important correction into account will give an erroneous velocity to the FDM particles. In turn this will result in numerical artifacts since the quantum pressure effects will be introduced once the numerical solver is activated. The FDM particles in the early stages of the simulation would then be ``pushed out" of proto-halos giving inaccurate mass accretion history for halos. In the case of fully FDM simulations, however, this would not be an issue since the displacement field of the FDM particles is already suppressed on small scales. This occurs since the transfer function for the whole system corresponds to that of FDM which is itself suppressed below the particle Jeans scale. By Eq.~(\ref{eq:scalar_poisson}), it follows that the velocity field will adopt the appropriate behaviour. Thus, problems arise once the transfer function of the whole system does not exactly describe FDM particles alone such as in the case of a low-concentration of ultralight particles in a mixed dark matter scenario. We have used the LPT calculator of the large-scale structure simulation  algorithm \texttt{Peak Patch} (\citealt{Stein2019TheValidation}) to generate the CDM velocity field of the bottom panel of Fig.~\ref{fig:disp_z80}. Although conceived to generate detailed halo catalogs, the code also allows the  user to create a density field realization from a power spectrum and output the corresponding LPT displacements as a function of redshift.

%%%%%%%%%%%%%%%%%%%%%%%%%%%%%%%%%%%
%%%%%%%%%%%%%%%%%%%%%%%%%%%%%%%%%%%

\subsection{Modified Deformation Tensor}\label{sec:deform_tensor}
Generally, two descriptions are used to model FDM particle the Schr\"odinger-Poisson system of Eq. (\ref{eq:Schro})-(\ref{eq:Poisson}) or the Madelung formalism Eq. (\ref{eq:Euler})-(\ref{eq:QP}). It is accepted that the first captures all of the features of the FDM physics, including the full extent of interference patterns in halos and filaments (\citealt{Mocz2019FirstFilaments}). The Madelung formalism can also lead to interference fringes (\citealt{Li2019NumericalModel}), but will not capture the full wave nature of FDM as the two representations are not equivalent unless quantization conditions are applied (\citealt{Wallstrom1994InequivalenceEquations}). The main issue is the fact that the phase $\theta$ is undefined at locations where the wavefunction vanishes. More specifically, what needs to be quantized is the line integral of the fluid velocity around any closed loop since this ensures the wavefunction is single-valued. This leads to the quantization of vortices forming in a superfluid governed by the Schr\"odinger-Poisson system (\citealt{Hui2020VorticesMatter}).

Unfortunately, evolving halos in the Schr\"odinger picture is quite computationally expensive and most of the simulations made are stopped at intermediate redshifts $z\sim 5$. Conversely, the Madelung equations can be more easily implemented in certain simulation algorithms as the only addition is that of the so-called quantum potential (see for example \citealt{Nori2018AX-GADGET:Models}). In these cases, the simulations can be run over much larger scales of order Mpc and and the stopping redshift gets put back further since the simulations take around 2-3 times the computational resources of the analogous CDM simulations (\citealt{Zhang2018UltralightSimulations}). The central point we wish to address here is that the the range (i.e. the spacetime locations) for which the Madelung formalism is valid is tightly linked to that of the Zeldovich approximation as both break down when close particle encounters occur. In other words, we can diagnose the failings of the Madelung picture by examining the times and locations where \emph{shell crossing} occurs (\citealt{Kopp2017SolvingMethod}). We will assess the validity of this statement by integrating the full Schr\"odinger-Poisson system in Section~\ref{sec:comparison}. In cases where the dark matter is mixed, we need to be more careful and isolate when the shell crossing occurs for the FDM particles only.

When using LPT, we can determine when and where shell crossing will happen through the deformation tensor of Eq.~(\ref{eq:deform_tens}). We've shown in our treatment of LPT that the displacement for the ultralight particles can be extracted from the usual displacements given a gravitational potential through the application of a low-pass filter. In other words,
\begin{align}
    \p^\mathrm{FDM}  = \mathcal{F}^{-1}\{L(k)\} \star \p^\mathrm{CDM},
\end{align}
where the $\star$ denotes convolution and $L(k)$ is defined in Eq.~(\ref{eq:lp}). By the properties of the convolution, we find that
\begin{align}
    \mathcal{D}_{ij}^\mathrm{FDM} = \delta_{ij} + \mathcal{F}^{-1}\{L(k)\} \star \frac{\partial \Psi^\mathrm{CDM}_i}{\partial q_j}.
\end{align}
By diagonalizing the above, we can find the evolution of overdensities
\begin{align}
    \delta(\q,t) = \frac{1}{\prod_i [1-D(t)\lambda_i(\q)]} -1, \label{eq:eigenvals}
\end{align}
where $\lambda_i$ are the eigenvalues of the deformation tensor at $\q$. We can determine if shell crossing occurred at a Lagrangian spacetime location $(\q,t)$ if the largest eigenvalue, say $\lambda_1$, satisfies 
\begin{align}
    D(t)\lambda_1(\q) > 1. \label{eq:SC-definition}
\end{align}
Using this approach, many mathematical studies have been devoted to the formation of caustics in LPT (see \citealt{Hidding2014TheComplexity,Feldbrugge2018CausticWeb} and references therein). Given the significant difference in computational costs between the two FDM modeling methods discussed above, the potential for constraints using computationally efficient $N$-body or hydrodynamical simulations will be dependent on whether one can use the Madelung formalism to describe mixed dark matter models accurately.

\begin{figure*}
    \centering
    \includegraphics[width=1.\linewidth]{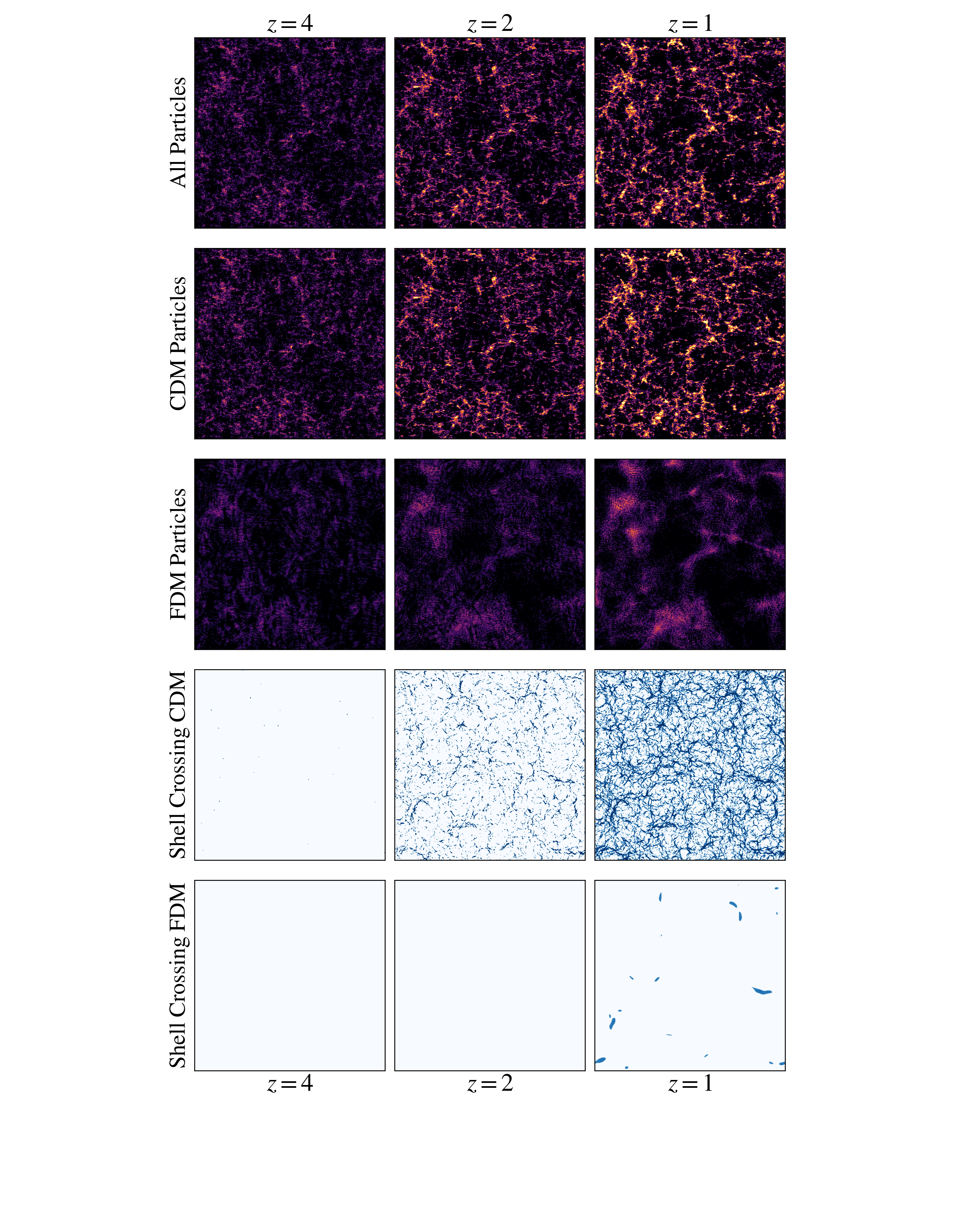}
    \caption{Simulated clustering using modified LPT (first order) and shell crossing regions: 
    (\textit{first row}) total dark matter distribution constituted from 10\% of $10^{-27}$ eV particles and 90\% CDM particles,
    (\textit{second row}) distribution of the CDM particles in the box,
    (\textit{third row}) distribution of the FDM particles in the box,
    (\textit{fourth row}) regions where shell-crossing occurred with CDM displacements in Lagrangian space,
    (\textit{fifth row}) regions where shell-crossing occurred with FDM displacements in Lagrangian space.
    }
    \label{fig:caustics}
\end{figure*}
\begin{figure}
    \centering
    \includegraphics[width=\linewidth]{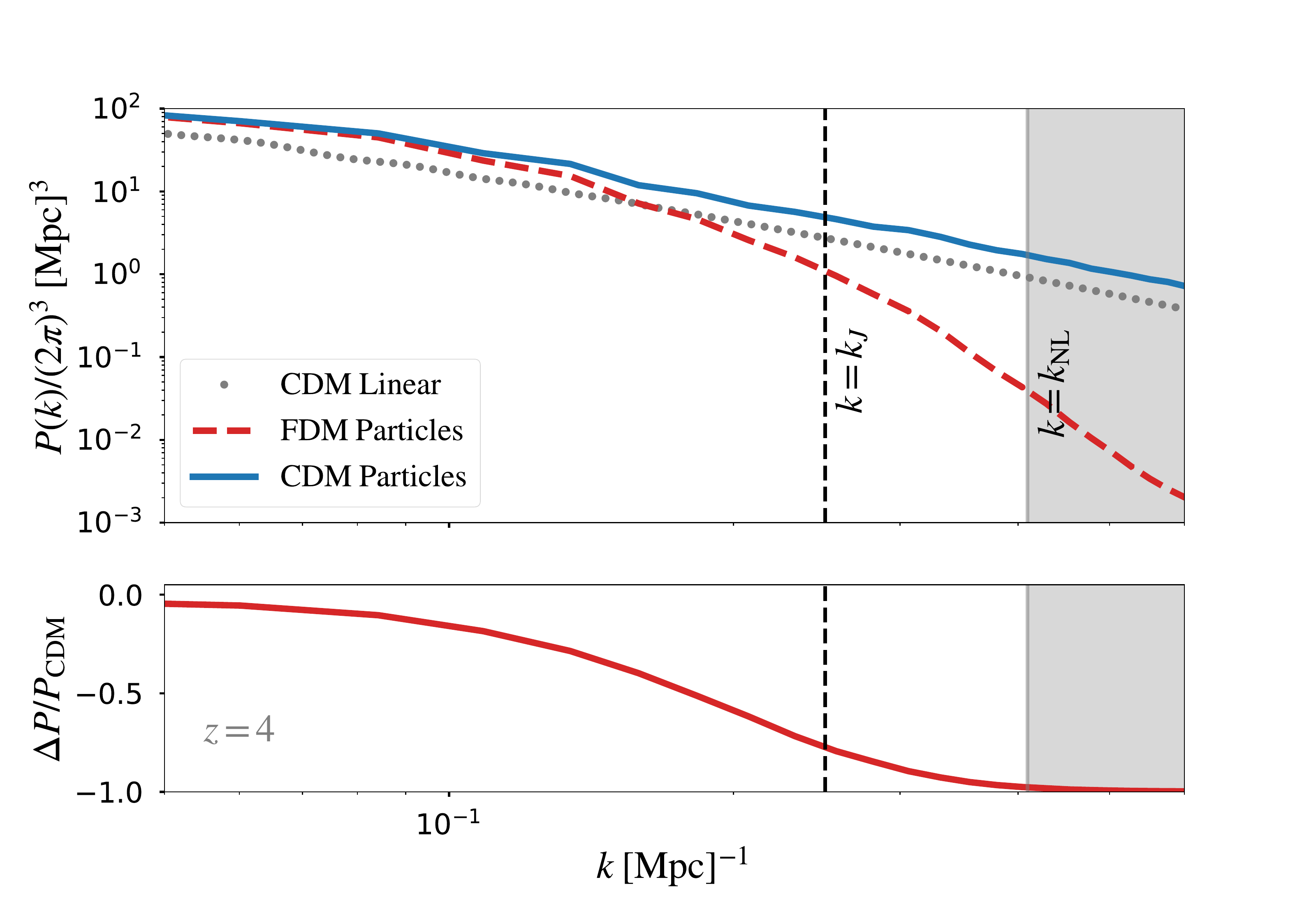}
    \caption{Power spectra of CDM and FDM particles part of a  90\%-10\% mixture inside a (256 Mpc)${}^3$ box at redshift $z=4$ subject to the same gravitational potential. These power spectra are that of panels 2 and 3 of the first column of Fig.~\ref{fig:caustics}. The shaded region corresponds to the non-linear regime. The linear prediction for CDM obtained from the \texttt{axionCAMB} code is plotted as well and is in good agreement with the CDM particle output at redshift $z=4$. %\renee{Can you plot the linear power spectra for both CDM and FDM in this case?I think this will be good to see the difference between the linear/non-linear and axion/cdm cases rather than mixing all the ideas using only two lines.}
    }
    \label{fig:pk_z4}
\end{figure}

To explore large-scale structure formation with the modified Zeldovich approximation at later times, we simulated the movement of test particles up to redshift $z=1$. We picked an illustrative DM configuration consisting of 10\% of $10^{-27}$ eV axions and 90\% CDM. In Fig.~\ref{fig:caustics}, we plotted the resulting structure at three snapshot redshifts for both the total, CDM, and FDM densities. We can see that despite the large proportion of CDM causing a steep potential well, the FDM particles cluster much later. This is reflected in the regions of shell-crossing of the two bottom rows of Fig.~\ref{fig:caustics}. A zoom-in version of the plot with second order effect and a FDM particle mass of $10^{-26}$ eV was presented earlier in Fig.~\ref{fig:mirror}. There, we also notice oscillatory patterns on top of the suppressed clustering. Note that pixelization effects arise additionally because of the limited resolution of LPT. Putting those aside, we find that generally, the size of the oscillations in FDM simulations are found to match the de Broglie wavelength of the particles (at first order).
This is due to the fact that suppressing high-frequency modes in a signal leads to perturbations known as ``ringing artifacts". In our case, these artifacts are part of our signal and not noise since the suppression of small-scale information comes from the physical nature of the DM and not from numerical effects. The frequency of these oscillating artifacts relate to the cutoff scale as $\lambda_\mathrm{osc} \sim 1/k_\mathrm{cutoff}$. It is also worth noting that the Jeans length $\lambda_J\sim 1/k_J \sim m^{-1/2}\rho^{-1/4}$ from Eq.~(\ref{eq:kJeans}). Following the reasoning of \cite{Marsh2016AxionCosmology} and taking the particle velocity to be $v^2\sim M/r$ with $M\sim \rho r^3$ then $\lambda_\mathrm{dB}\sim 1/mv \sim 1/m\rho^{1/2}r$. Considering scales close to the de Broglie wavelength, we can set $\lambda_{dB} \sim r$ which gives $\lambda_{dB} \sim m^{-1/2}\rho^{-1/4}$. Finally, we arrive at the conclusion that the observed oscillations follow
\begin{align}
    k_\mathrm{cutoff} \sim k_J \Rightarrow \lambda_\mathrm{osc} \sim \lambda_J \sim \lambda_\mathrm{dB},
\end{align}
as claimed.

From the two last rows of Fig.~\ref{fig:caustics} , we observe that shell-crossing happens much sooner and to a much larger extent for CDM than it does for FDM particles. This is very similar to what was found for the truncated Zeldovich approximation where the small-scale suppression of displacements is put in as a correction to prevent shell-crossing rather than as a physical effect (\citealt{Melott1994OptimizingApproximation}). Note that we do not expect the regions of shell-crossing to match the overdense regions of the three top rows since shell-crossing occurs when the largest eigenvalue of the deformation tensor satisfies Eq.~(\ref{eq:SC-definition}) which is expressed in Lagrangian coordinates ($\q$) while the final particle positions are functions of the Eulerian coordinates ($\x$). In order to investigate the grid-like patterns of the FDM particles at redshift 4, we calculate the power spectra of the particles inside the box. The results are plotted in Fig.~\ref{fig:pk_z4} along with the Jeans scale $k_J$ and the scale at which non-linearities begin to be of importance which we can find by imposing $k^3P(k_\mathrm{NL},z)/(2\pi)^2 = 1$. We notice the familiar power spectrum suppression characteristic of the FDM clustering and we find that it matches the imposed scale-dependence by showing approximately a 75-80\% reduction in clustering at $k=k_J$.

\subsection{Comparison to other simulation techniques}\label{sec:comparison}
We have advertised two main uses for the modified LPT which are to include the effects of the quantum pressure term in the particle velocities for initial conditions and to use the occurrence of shell-crossing as a diagnostic tool. We wish to compare the modified LPT scheme with the two other main simulation approaches for FDM which are the Schr\"odinger wavefunction and the Madelung approximation. Given a density field, we can calculate the quantum pressure arising in the Madelung approximation from Eq.~(\ref{eq:QP}). We compare the value of the quantum pressure to the displacement suppression between CDM and FDM using the fact that at linear order Eq.~(\ref{eq:Euler}) can be expressed as
\begin{align}
    \mathbf{v}_\mathrm{FDM} \sim -\frac{1}{a}\nabla (\Phi + Q).
\end{align}
Therefore, we expect (with the use of Poisson's equation)
\begin{align}
    \frac{1}{a}\nabla \cdot \mathbf{v}_\mathrm{FDM} \sim -4\pi G\bar{\rho}\delta -\frac{1}{a^2}\nabla^2 Q,
\end{align}
where the factors of $a^{-1}$ were absorbed in the Poisson equation when transforming the Laplacian from comoving to physical coordinates. We can also linearize the quantum potential by considering small local deviations in the density with respect to the cosmological average. Taking the Fourier transform of the result then gives the first order expression (\citealt{Marsh2015NonlinearForces})
\begin{align}
    \mathcal{F}\{Q\} \approx -\frac{\hbar^2k^2}{4m^2a^2}\mathcal{F}\{\delta\}.
\end{align}
The full expression in Fourier space becomes
\begin{align}
    -\frac{i\mathbf{k}}{a}\cdot \mathcal{F}\{\mathbf{v}_\mathrm{FDM}\} \sim -\Big(4\pi G\bar{\rho}-\frac{\hbar^2k^4}{4m^2a^4}\Big)\mathcal{F}\{\delta\},
\end{align}
which matches exactly the two last terms of Eq.~(\ref{eq:PDE_growth}). We thus expect the modified LPT scheme to recover the physics of the Madelung formalism in the regime where $\delta\ll 1$. The modified LPT approach presented here does not recover the full oscillatory behaviour of FDM on small scales after shell-crossing has occurred. This limitation is shared by the Madelung approximation, but not the Schr\"odinger picture. This makes it difficult to compare the two approaches \textit{a priori}. This is when we can turn to the deformation tensor of Sec.~\ref{sec:deform_tensor}. The prediction from the adapted deformation tensor was that light particles should experience shell-crossing later than predicted by the usual Zeldovich approximation for CDM. This is particularly useful when considering the breakdown of the Madelung approximation at the formation of interference fringes (i.e. at points where the density vanishes).

\begin{figure}
    \centering
    \includegraphics[width=\linewidth]{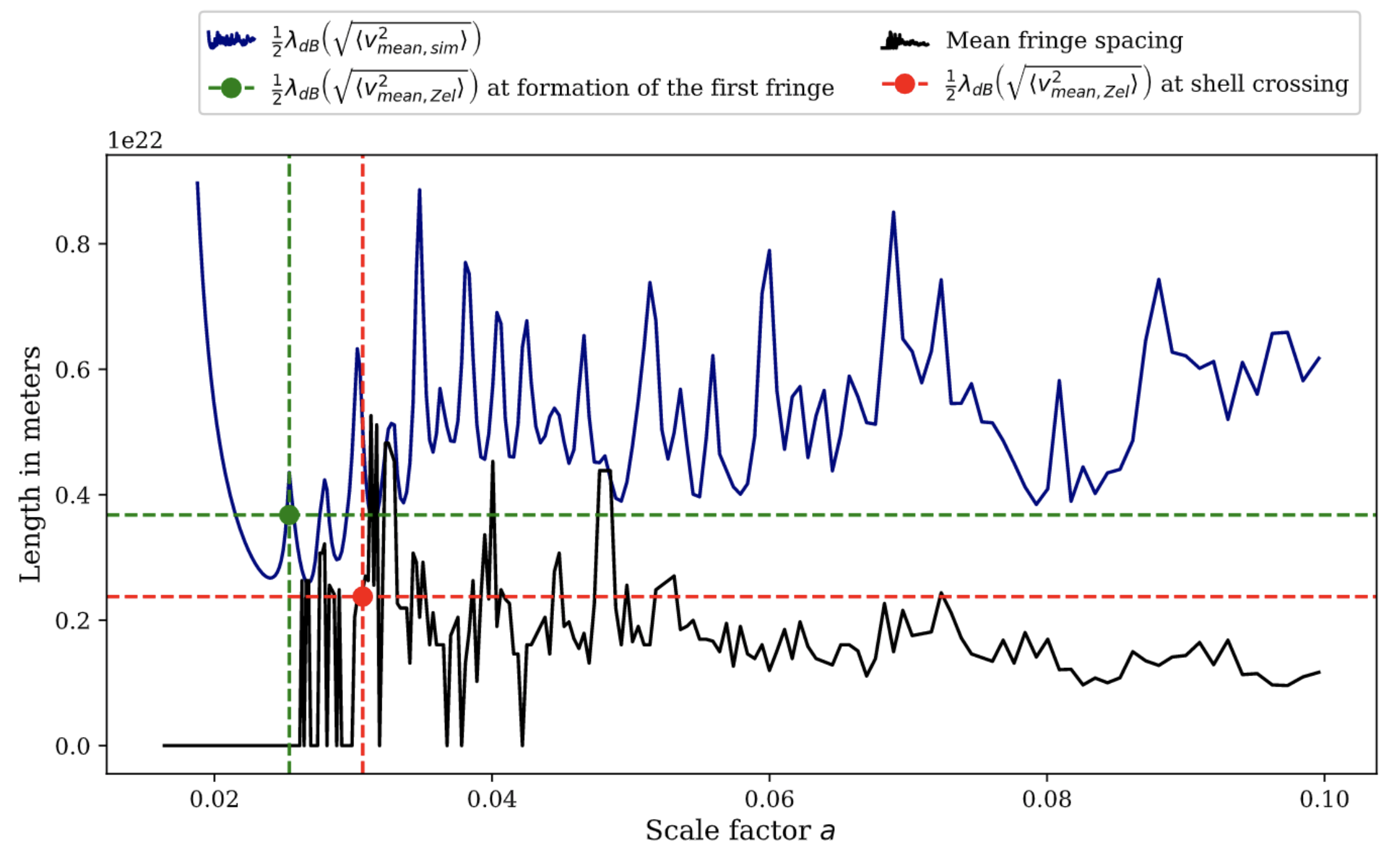}
    \caption{Interference fringe spacing in cylindrical collapse (black line). The mean fringe spacing is well approximated  by a half de Broglie wavelength computed from the mean square Zeldovich velocity at shell crossing (red dashed line). This is a better approximation than using the locally measured mean square velocity (blue line), or the Zeldovich velocity at the time the first fringe forms (green dashed line). The scale factor time coordinate is given arbitrary normalisation.}
    \label{fig:zeldovich_sch2}
\end{figure}

We tested the (CDM) Zeldovich approximation against a numerical solution of cylindrical collapse for the Schr\"odinger-Poisson equations, Eq.~(\ref{eq:Schro})-(\ref{eq:Poisson})\footnote{We used the simulation code PyUltralight (\citealt{Edwards2018PyUltraLight:Dynamics})}, in order to explore a toy model for FDM in galactic filaments.
\begin{figure}
    \centering
    \includegraphics[width=\linewidth]{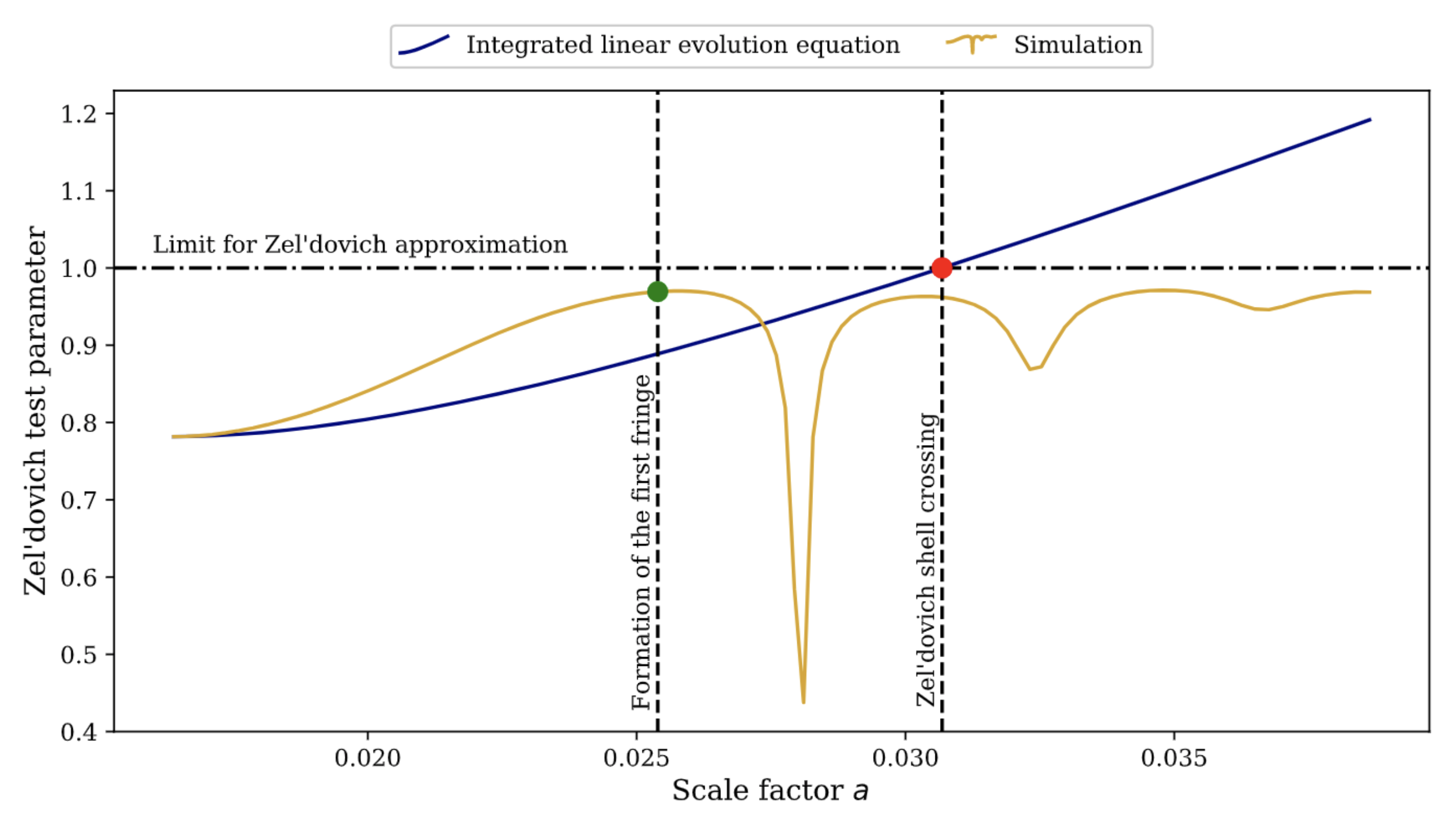}
    \caption{Value of the Zeldovich test parameter as a function of time in a PYUltralight simulation of a collapsing filament. The blue line depicts the usual linear CDM case while the yellow line describes the full evolution of the wavefunction.}
    \label{fig:zeldovich_sch}
\end{figure}
%%%%%%%
The cylindrical collapse simulation provides insights about the formation of interference fringes in FDM. The first fringe forms shortly before shell crossing, and is identified by locating discontinuities in the velocity field and phase of the wave function. Such discontinuities allow the fringe spacing to extracted from the simulation. We first examine the closeness to shell-crossing in the collapsing filament. For this we define the Zeldovich test parameter through the relation
\begin{align}
    D(a)\lambda_1 = 1 - \bigg(\frac{\bar{\rho}}{\rho}\bigg)^{1/2}, \label{eq:test_param}
\end{align}
where $\lambda_1$ is the largest eigenvalue of the deformation tensor of Eq.~(\ref{eq:deform_tens}) (note that in the case of a filament structure we can take $\lambda_3 =0$ and $\lambda_2=\lambda_1$ by symmetry) and where $\bar{\rho}$ is the mean density in the box. The the LHS of Eq.~(\ref{eq:test_param}) is then obtained from the Zeldovich approximation while the RHS is taken directly from the wavefunction being evolved in the simulation and is defined as the test parameter. Fig.~\ref{fig:zeldovich_sch2} shows the fringe spacing as a function of time during collapse, which is well approximated by a half of de Broglie wavelength with $v$ computed from the mean square Zeldovich velocity at shell crossing. 

The presence of interference fringes in FDM galactic filaments is a striking and unique prediction of the model~\cite{Schive2014CosmicWave,Mocz2019FirstFilaments}. Furthermore, there are no pressure supported stable cores in cosmic FDM filaments (like it's the case in 3D structure forming soliton cores) which means filaments develop a range of interesting substructures not generally present in halos (\citealt{Mocz2019FirstFilaments}). The utility of the Zeldovich approximation for predicting the fringe properties without recourse to full Schr\"odinger-Poisson simulations could be of use in formulating observational searches for FDM interference.
%%%%%%
%%%%%%%

%%%%%%%%%%%%%%%%%%%%%%%%%%%%%%%%%%%
%%%%%%%%%%%%%%%%%%%%%%%%%%%%%%%%%%%

\section{Discussion \& Conclusions} \label{sec:conclusion}

We have adapted LPT to describe the scale-dependent pressure arising with ultralight (fuzzy) dark matter particles. We have smoothed the fast oscillations of the linear growth factor using a modified Heaviside step-function model and we have found the scaling solutions for its parameters as functions of particle mass and redshift. We also generalized the expression for the smoothed axion growth factor to mixed DM models and developed a normalization which does not lead to divergences for large $k$. We then used the formalism developed in \cite{Tatekawa2002PerturbationDispersion} to implement the modified growth factor into LPT. We adapted the outputs of LPT calculations to obtain a description of the flow of CDM and FDM given a common initial density field. We made use the code \texttt{Peak Patch} to solve for the the LPT evolution equations in the CDM case. From there, we considered universes composed of a mixture of 10\% FDM particles of mass $m=10^{-26}$ eV (Fig.~\ref{fig:mirror}) or $m=10^{-27}$ eV (Fig.~\ref{fig:disp_z80}-\ref{fig:caustics}) with the remaining dark matter behaving as CDM. It is worth noting that the scheme developed here is extendable to an arbitrary number of particles and can model the $\mathcal{O}(100)$ light axion scenarios suggested by string theory.

Since one of the primary uses of LPT is to generate initial conditions for $N$-body simulations, we calculated the effects of the modified growth on the initial velocities of ultralight particles at high redshift. The magnitude and direction of the trajectories can vary by a significant amount in regions of high density and at high redshift as shown in Fig.~\ref{fig:disp_z80}. We conclude that in the event where $N$-body simulations are run with mixed dark matter, one would need to attribute the particles the corresponding velocities as a function of their mass. This is necessary in order for the initial density and velocity fields to be consistent with the dynamical evolution which accounts for the quantum pressure effects.

Finally, we also considered lower redshift implications of our findings for large-scale structure formation in mixed DM universes. We simulated the evolution of the cosmic web at different redshifts even after shell-crossing to investigate two phenomena. The first was the response of the ultralight particles to the increase in the gravitational potential on small-scales (as compared with the pure FDM case for which the potential is suppressed). We found that the particles do eventually fall into the potential well created by the CDM, but still exhibit a lack of clustering on small scales due to quantum unceratainty. From more advanced simulations, one could potentially differentiate between pure CDM and mixed DM clustering using probes such as gravitational lensing. For the second phenomenon, we studied the impact of the modified growth factor on the occurrence of shell-crossing. We found that shell-crossing was delayed for the ultralight particles compared to CDM. We also used the code PyUltralight to solve the Schr\"odinger-Poisson system for a collapsing filament. We concluded that the appearance of shell-crossing does correspond closely to the time of formation of interference patterns. Thus shell-crossing in modified LPT can be used as a diagnostic tool to determine the temporal range of validity for the Madelung approximation (which breaks down when interference fringes first form). We also showed that LPT velocity can be used to predict the scale of oscillations and interference fringes in FDM structure.

Although more reliable than other simulation techniques on small-scales and at late times, the main impediment of the Schr\"odinger approach has to do with the timestep size necessary to simulate the evolution of the system accurately. First, one must resolve the de Broglie wavelength of the particles even in the case where the density is slowly evolving since the wavefunction phase still changes. This is particularly difficult as the wavelength shrinks as the particles gain velocity falling into gravitational potential wells. In comoving boxes, this problem is exacerbated by the fact that the grid on which the Schr\"odinger-Poisson system is solved simultaneously increases with the scale factor. On the other hand, for the particles on the lighter end it seems possible to escape these restraints. From the CFL condition for parabolic systems, we have that $\Delta t\leq \min\Big[(\Delta x)^2 ma^2 / (6\hbar), \hbar/m\Phi\Big]\leq \lambda_\mathrm{dB}^2 ma^2 / (6\hbar)$ (\citealt{Li2019NumericalModel}). This implies that the upper bound on the timestep will grow as $1/m$ since $\lambda_\mathrm{dB}\sim 1/m$ and thus the computational cost of mixed DM models may be much lower than those for full FDM simulations if one considers low-concentration, low-mass scenarios. Also, a modified LPT taking the dynamical effects into account could allow for simulations to be started at a lower redshift. Therefore, despite an increase in parameter-space configurations to cover, mixed ultralight DM models can still be pursued at a reasonable computational cost given that both the Madelung and Schr\"odinger treatments are efficient modeling techniques for a wide range of scenarios.

%%%%%%%%%%%%%%%%%%%%%%%%%%%%%%%%%%%
%%%%%%%%%%%%%%%%%%%%%%%%%%%%%%%%%%%

\section*{Acknowledgments}
We would like to thank Phil Mocz for useful comments and suggestions. RH is a CIFAR Azrieli Global Scholar, Gravity \& the Extreme Universe Program, 2019, and a 2020 Alfred. P. Sloan Research Fellow. RH is supported by Natural Sciences and Engineering Research Council of Canada. RB is a CIFAR Fellow. AL, RB and RH are supported by Natural Sciences and Engineering Research Council of Canada. The Dunlap Institute is funded through an endowment established by the David Dunlap family and the University of Toronto. DJEM is supported by the Alexander von Humboldt Foundation and the German federal Ministry of Education and Research, and acknowledges the hospitality of King's College London, where part of this work was completed. LS thanks David Ellis for valuable discussions. We acknowledge that the land on which the University of Toronto is built is the traditional territory of the Haudenosaunee, and most recently, the territory of the Mississaugas of the New Credit First Nation. We are grateful to have the opportunity to work in the community, on this territory.

%%%%%%%%%%%%%%%%%%%%%%%%%%%%%%%%%%%
%%%%%%%%%%%%%%%%%%%%%%%%%%%%%%%%%%%

\bibliographystyle{plainnat}
\bibliography{references}

%%%%%%%%%%%%%%%%%%%%%%%%%%%%%%%%%%%
%%%%%%%%%%%%%%%%%%%%%%%%%%%%%%%%%%%
\newpage
\appendix

\section{On the Effects of Smoothing}\label{sec:app_osc}
Throughout this work, we have neglected the small oscillations of the growth factor for $k>k_J$ and averaged it out to zero. In this section, we explore the consequences of keeping the oscillations and propagating them using first order LPT to the displacements. To simplify things, we consider a one-dimensional model consisting of a single Gaussian overdensity. To isolate the effects of smoothing (averaging) the oscillations, we use three configurations for the growth factor: (1) the usual CDM growth, (2) the smoothed FDM growth, and (3) the full growth FDM with oscillations.
\begin{figure}
    \centering
    \includegraphics[width=1.\linewidth]{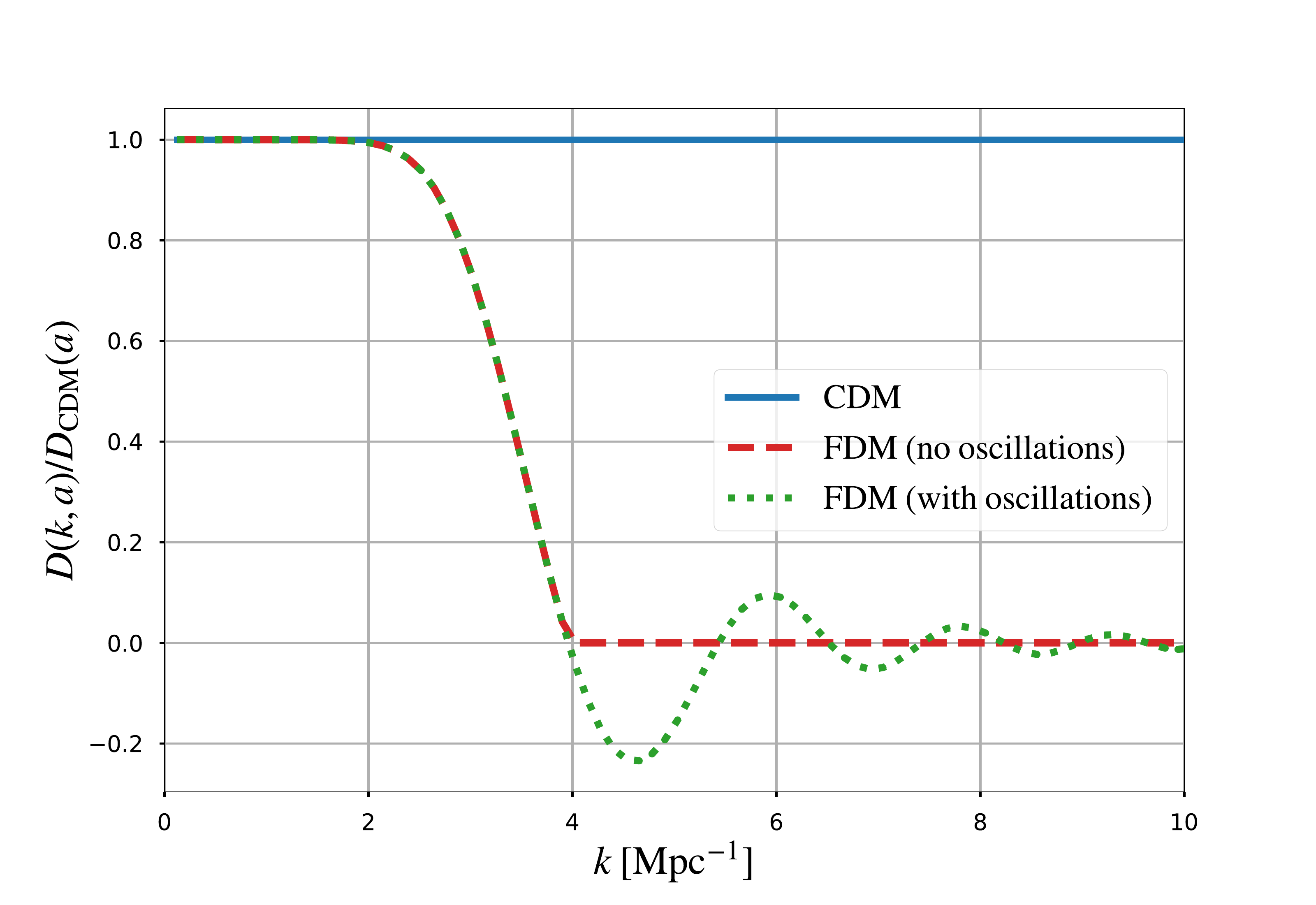}
    \caption{FDM linear growth factor considering small-scale oscillatory behaviour or setting oscillations to zero. The CDM growth is scale independent. The example here is that of a growth factor with mass $m\approx 4\times 10^{-23}$ eV, with FDM fraction $F=1$ at redshift $z=50$.}
    \label{fig:growth_osc_smooth}
\end{figure}
\begin{figure}
    \centering
    \includegraphics[width=1.\linewidth]{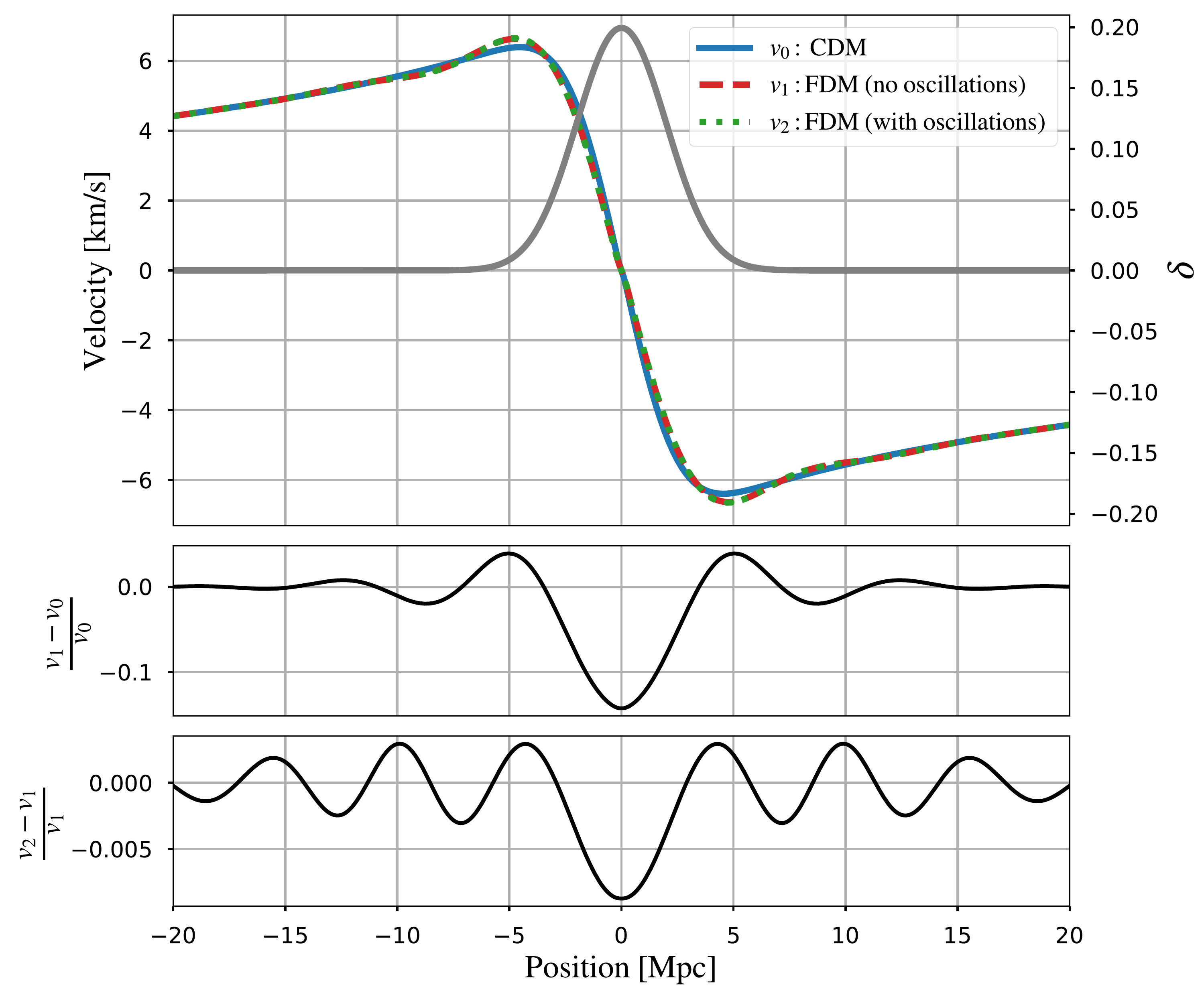}
    \caption{(\textit{Top}) First order LPT velocity resulting from different growth factors presented in Fig.~\ref{fig:growth_osc_smooth}. The central Gaussian overdensity creating the gravitational potential is shown in grey. The velocity curves are positive when the 1D velocity vector is pointing to the right and negative when it points to the left. (\textit{Middle}) Relative difference between CDM and FDM velocities. (\textit{Bottom}) Relative difference between the two FDM approaches to examine the impact of suppressing the small scale oscillations in the growth factor.}
    \label{fig:disp_osc_smooth}
\end{figure}
The three cases are presented in Fig.~\ref{fig:growth_osc_smooth} where the FDM growth factor is obtained using the renormalized Bessel function approach of Eq.~(\ref{eq:li_grow})-(\ref{eq:taylor_exp}) and where the averaging is simply done by setting all values after the first node to zero. Note that the case plotted corresponds to an FDM fraction of $F=1$. From the modified growth, we calculate the LPT displacements at first order and compared the CDM results to the FDM as well as the two different FDM approaches between themselves. The results are shown in Fig.~\ref{fig:disp_osc_smooth}. We observe that the introduction of the oscillations on small scales causes a ringing effect in the particle velocity. By increasing the value of the overdensity $\delta$ and by reducing its variance, we can increase the magnitude of this effect to a percent-level correction. We therefore conclude that we can safely neglect the impact of these oscillations for $\delta\ll 1$.
\begin{figure}
    \centering
    \includegraphics[width=1.\linewidth]{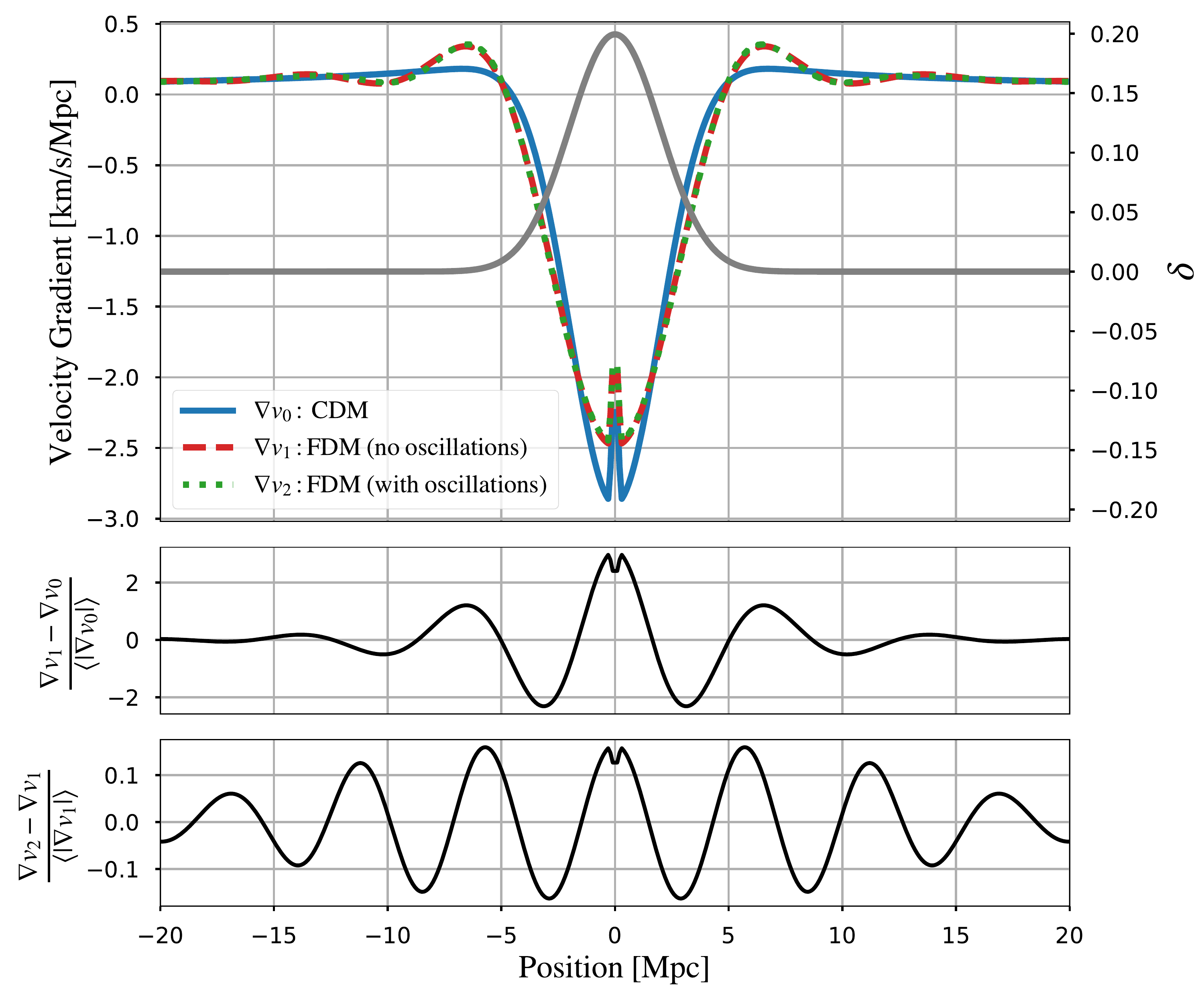}
    \caption{Gradient of velocities shown in Fig.~\ref{fig:growth_osc_smooth} with relative difference between gradients. The relative difference is taken as the mean of the magnitude since the oscillating growth as many points of null gradients.} 
    \label{fig:grad_disp_osc_smooth}
\end{figure}
Derivatives of displacements are more sensitive to small-scale effects and comparing gradients (see Fig.~\ref{fig:grad_disp_osc_smooth}) reveals some periodic patterns of higher amplitude due to the presence of oscillations in the growth factor. The presence of these oscillations are a unique signature of FDM, but we leave the treatment of higher order derivatives of the displacement fields for future work.

\section{Second Order Derivation}\label{sec:appendix-2lpt}
In this work, we have claimed that the algorithm developed to obtain initial displacements and velocities can be extended to second order Lagrangian perturbation theory (2LPT). Since most initial condition calculating codes work at second order (e.g. \texttt{MUSIC} \citealt{Hahn2011Multi-scaleSimulations}), we justify this claim in more detail. The proof consists of the following demonstrations:
\begin{enumerate}
    \item show the large scale behaviour of 2LPT displacements matches CDM,
    \item show the ``transition region" where the growth isn't fully suppressed is small,
    \item show the small-scale behaviour of 2LPT growth is given by the convolution of a decaying Green function and a bounded source term,
    \item conclude the small-scale displacements decay quickly to zero on small scales.
\end{enumerate}

\subsection{High Redshift Approximations}
In order to justify applying our small-scale smoothing procedure to second order, the first observation we will make is that the region of $k$-space where the growth is neither suppressed or perfectly following CDM is very small at high redshift. To see this consider a small positive $0<\epsilon<1$ and let
\begin{align}
    \Delta_\epsilon = k_+(\epsilon) - k_-(\epsilon),
\end{align}
where $k_\pm(\epsilon)$ are defined such that $L(k_+)=1-\epsilon$ and $L(k_-)=\epsilon$. Then, we can invert the low-pass filter to obtain
\begin{align}
    \Delta_\epsilon = \frac{1}{2\alpha}\bigg[\ln\bigg(\frac{1}{\epsilon^{1/8}}-1\bigg) - \ln\bigg(\frac{1}{(1-\epsilon)^{1/8}}-1\bigg) \bigg].
\end{align}
For a value of $\epsilon = 0.05$, we have that
\begin{align}
    \Delta_{0.05} \approx \frac{1}{\alpha} \lesssim \mathcal{O}(1)\;\mathrm{Mpc}^{-1}, \label{eq:small_transition}
\end{align}
for $m\lesssim 10^{-24}$ eV. So from Eq.~(\ref{eq:small_transition}), we have that the decay is extremely rapid and the transition region is small. Therefore the behaviour of the growth is determined solely by the asymptotic behaviour on very large and very small scales. Thus to demonstrate that the current approximation for the scale-dependence of the growth factor is valid at second order, it suffices to show that we expect the second order displacements to match CDM on large scales and to be zero on small scales (above the Jeans scale). On large scales, this is immediate since even before perturbatively expanding the equation for the displacements the influence of the quantum pressure is only noticeable when $k>k_J$. This follows from the definition of the axion Jeans scale. It remains to show that the second order displacements vanish below the Jeans scale, which we will do in the upcoming section.

%%%%%%%%%%%%%%%%%%%%%%%%%%%%%%%%%%%

\subsection{Small-Scale Evolution}
Building on the impressive work of \cite{Morita2001ExtendingDispersion}, we claim the procedure developed to map the CDM displacement field to the FDM displacement as a convolution with a low-pass filter applies to second order LPT as well. We go back to Eq.~(\ref{eq:traj}) and rewrite it in terms of the displacement potential $\phi$ and Lagrangian coordinates $\q$. Following the same reasoning as before, we will expand the displacement potential such that $\p^{(1,2)}=\nabla_
\q\phi^{(1,2)}$. This gives at first order
\begin{align}
    \nabla_\q^2 \bigg(\frac{d^2\phi^{(1)}}{d\tau^2} + 2 \mathcal{H} \frac{d\phi^{(1)}}{d\tau} - \frac{3}{2}\mathcal{H}^2\Omega_m\phi^{(1)} - \frac{1}{a^2}\frac{dP}{d\rho}\nabla^2_\q\phi^{(1)}\bigg) = 0.
\end{align}
With a reasonable choice of boundary conditions, we have (\citealt{Buchert1992LagrangianApproximation})
\begin{align}
    \frac{d^2\phi^{(1)}}{d\tau^2} + 2 \mathcal{H} \frac{d\phi^{(1)}}{d\tau} - \frac{3}{2}\mathcal{H}^2\Omega_m\phi^{(1)} - \frac{1}{a^2}\frac{dP}{d\rho}\nabla^2_\q\phi^{(1)} = 0.
\end{align}
To first order, we could solve the spatial and temporal and growth parts of the equation of motion separately and then express the scale dependence of the growth along with the CDM solution. At second order, there is mode coupling and this is no longer possible. However, we can solve the equations for the second order displacement and study its total scale dependence to see if it matches what we observed at first order. To this end, we express the second order equation of motion as
\begin{align}
    \nabla_\q^2 \bigg(\frac{d^2\phi^{(2)}}{d\tau^2} &+ 2 \mathcal{H} \frac{d\phi^{(2)}}{d\tau} \nonumber\\&- \frac{3}{2}\mathcal{H}^2\Omega_m\phi^{(2)} - \frac{1}{a^2}\frac{dP}{d\rho}\nabla^2_\q\phi^{(2)}\bigg) = \mathcal{Q}(\q,\tau),
\end{align}
where $\mathcal{Q}(\q,\tau)$ (not to be confused with the quantum pressure) is the source term which is defined as
\begin{align}
    \mathcal{Q}(\q,\tau) \equiv &2\pi G\bar{\rho} \Big[\phi^{(1)}_{,ij}\phi^{(1)}_{,ij}-\Big(\nabla_\q\phi^{(1)}\Big)^2 \Big] \nonumber
    \\&-\frac{1}{a^2}\frac{dP}{d\rho} \Big[\nabla_\q\phi_{,i}^{(1)}\nabla_\q\phi_{,i}^{(1)}+\phi^{(1)}_{,ijk}\phi^{(1)}_{,ijk}+2\phi_{,ij}^{(1)}\nabla_\q\phi^{(1)}_{,ij}\Big] \nonumber
    \\&-\frac{1}{a^2}\frac{d^2P}{d\rho^2} \bar{\rho} \Big[\nabla^2_\q\phi^{(1)}\nabla^2_\q\nabla^2_\q\phi^{(1)}+\nabla^2_\q\phi_{,i}^{(1)}\nabla^2_\q\phi_{,i}^{(1)} \Big], \label{eq:second-order-phi}
\end{align}
where the comma denotes the partial derivative with respect to the Lagrangian coordinate ($\partial_{q_i}$) and where repeated indexes are implicitly summed over. Note here that to be consistent with the rest of our approach, we assume an irrotational displacement field and remove all transverse modes. Taking the Fourier transform of Eq.~(\ref{eq:second-order-phi}), we can express the solution for the second order displacement potential as an integral with a Green function $G(k,a,a^\prime)$ giving
\begin{align}
    \mathcal{F}\{\phi^{(2)}\} = -\frac{1}{k^2} \int^a da^\prime G(k,a,a^\prime) \mathcal{F}\{\mathcal{Q}\}(\mathbf{k},a^\prime). \label{eq:2lpt-main-integral}
\end{align}
The Fourier transform of the source reads
\begin{align}
    \mathcal{F}\{\mathcal{Q}\}(\mathbf{k},a^\prime) \propto \int_{-\infty}^\infty d^3\mathbf{k}^\prime &\mathcal{F}\{\phi^{(1)}\}(\mathbf{k}^\prime,t) \mathcal{F}\{\phi^{(1)}\}(\mathbf{k}-\mathbf{k}^\prime,t)\nonumber \\& \times U\bigg(\mathbf{k}-\mathbf{k}^\prime,\mathbf{k}^\prime,\frac{dP}{d\rho},\frac{d^2P}{d\rho^2}\bigg),
\end{align}
where the function $U$ is defined in \cite{Morita2001ExtendingDispersion}. The key aspect of interest is that it is composed of multiple products of wavenumbers such that it can be written as a two-variable polynomial in the norms $k$ and $k^\prime$. Also, since $\mathcal{F}\{\phi^{(1)}\}$ decays immediately after the axion Jeans scale $k_J$, we have that
\begin{align}
    &\int_{-\infty}^\infty d^3\mathbf{k}^\prime \mathcal{F}\{\phi^{(1)}\}(\mathbf{k}^\prime,t) \mathcal{F}\{\phi^{(1)}\}(\mathbf{k}-\mathbf{k}^\prime,t) U \nonumber
    \\&\lesssim \int_{-k_J}^{k_J} d^3\mathbf{k}^\prime \mathcal{F}\{\phi^{(1)}\}(\mathbf{k}^\prime,t) \mathcal{F}\{\phi^{(1)}\}(\mathbf{k}-\mathbf{k}^\prime,t) U.
\end{align}
From here, we deduce that the integrand of Eq.~(\ref{eq:2lpt-main-integral}) is bounded for all wavenumbers $k$ if the second derivative of the pressure is also bounded. This is immediate in the small-scale regime since the sound speed is given \emph{exactly} by (\citealt{Marsh2016AxionCosmology})
\begin{align}
    c_s^2 = \frac{\hbar^2k^2/4m^2a^2}{1+\hbar^2k^2/4m^2a^2},
\end{align}
where the approximate form of Eq.~(\ref{eq:soundspeed}) holds for $k\lesssim k_J$ which is where we've been computing our values until this point. However, to make sure the integral is bounded, we must ensure that this holds as $k\to \infty$. What we notice is that
\begin{align}
     \lim_{k\to \infty} \frac{dP}{d\rho} = \lim_{k\to \infty} c_s^2 = 1 \Rightarrow \lim_{k\to \infty} \frac{d^2P}{d\rho^2} = 0.
\end{align}
From this we can conclude that the source term $\mathcal{F}\{Q\}$ does not diverge in Fourier space.

Going back to the main expression to solve of Eq.~(\ref{eq:2lpt-main-integral}), we can focus on the first component of the convolution we have that the Green function is given by
\begin{align}
    G(k,a,a^\prime) = -&\frac{\pi}{2\sin(\nu\pi)} \bigg(\frac{4}{3}-\frac{5}{6\nu}\bigg)^{-1} a^{-1/4} a^{\prime 7/4}\mathcal{A}(k,a,a^\prime),
\end{align}
where
\begin{align}
    \mathcal{A}(k,a,a^\prime)\equiv &C_\pm(k) \bigg[J_{-\nu}\Big(\hbar k^2/mH_0\sqrt{a}\Big)J_{\nu}\Big(\hbar k^2/mH_0\sqrt{a}\Big)\nonumber\nonumber \\&-J_{\nu}\Big(\hbar k^2/mH_0\sqrt{a}\Big)J_{-\nu}\Big(\hbar k^2/mH_0\sqrt{a}\Big) \bigg],
\end{align}
and where $C_\pm(k)= C_+(k)C_-(k)$ is the appropriate normalization constant for the growing and decaying modes (previously we only considered the growing solution). Using the fact that the Jeans scale $k_J=a^{1/4}\sqrt{mH_0}$, we can rewrite $\mathcal{A}$ as
\begin{align}
    \mathcal{A}(k,a,a^\prime) = &C_\pm(k) \times\nonumber\\& \Bigg[J_{-\nu}\Bigg(\frac{k^2}{k_J^2}\Bigg) J_{\nu}\Bigg(\frac{k^2}{k_J^2}\sqrt{\frac{a}{a^\prime}}\Bigg) -J_{\nu}\Bigg(\frac{k^2}{k_J^2}\Bigg) J_{-\nu}\Bigg(\frac{k^2}{k_J^2}\sqrt{\frac{a}{a^\prime}}\Bigg)\Bigg]. \label{eq:A-function}
\end{align}
\begin{figure}
    \centering
    \includegraphics[width=\linewidth]{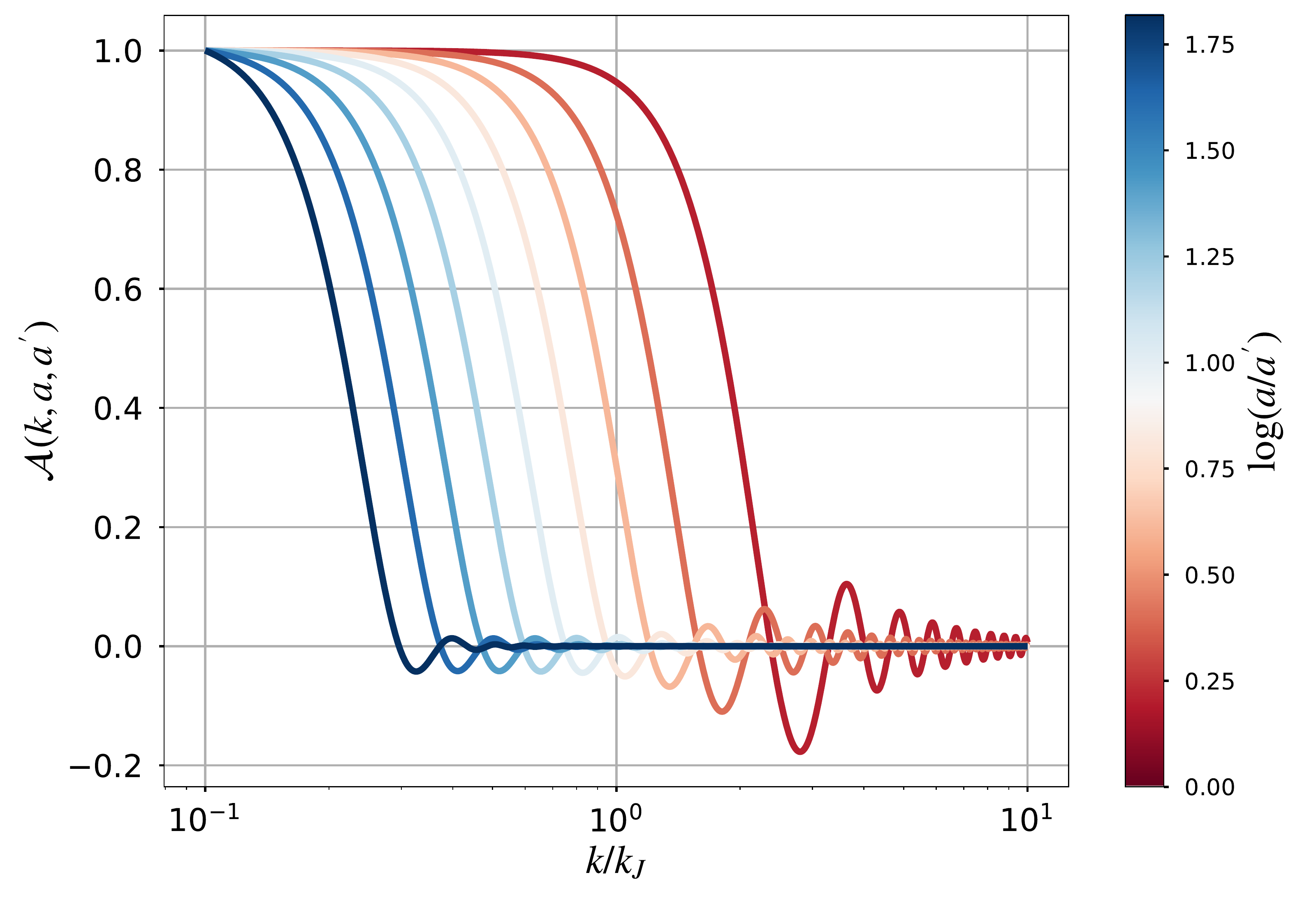}
    \caption{Plot of the function $\mathcal{A}$ defined in Eq.~(\ref{eq:A-function}) which encapsulates the scale-dependence of the Green function $G$. We observe a sharp drop for $k>k_J$ as was observed for the growth factor at first order. Due to the integration range of Eq.~(\ref{eq:2lpt-main-integral}), the value of $a^\prime$ never exceeds the value of $a$.}
    \label{fig:second-order-decay}
\end{figure}
Plotting $\mathcal{A}$ for different values shown in Fig.~\ref{fig:second-order-decay} of $a^\prime<a$, we observe a quick decay for $k>k_J$. Given that all other quantities involved in Eq.~(\ref{eq:2lpt-main-integral}) are bounded, we conclude that
\begin{align}
    \mathcal{F}\{\phi^{(2)}\} \approx 0 \;\;\;\;\forall k>k_J,
\end{align}
as claimed.

%%%%%%%%%%%%%%%%%%%%%%%%%%%%%%%%%%%
%%%%%%%%%%%%%%%%%%%%%%%%%%%%%%%%%%%

\bsp	% typesetting comment
\label{lastpage}
\end{document}